\documentclass[12pt,twoside]{article}
\usepackage[utf8]{inputenc}
\usepackage[backend=bibtex,style=numeric-comp,autocite=plain,sorting=none,%
giveninits=true]{biblatex}
\usepackage{amsmath,amsfonts,amssymb,amsthm,mathabx}
\usepackage{times}
\usepackage{hyperref}
\hypersetup{
  breaklinks=true,   %
  colorlinks=true,   %
}
\usepackage{xurl}
\usepackage{tensor}
\usepackage{color}
\usepackage{caption}
\usepackage{enumitem}
\usepackage{array}
\usepackage{amsbsy}
\usepackage{appendix}
\usepackage{mathrsfs}
\usepackage{physics}
\usepackage{cancel}
\usepackage{yfonts}
\usepackage{chngcntr}
\usepackage{xcolor}
\usepackage{soul}
\usepackage{cancel}
\usepackage{subfigure}
\usepackage{wrapfig}
\usepackage[leqno,fleqn,intlimits,newmultline]{empheq}
\usepackage{enumitem}

\usepackage{graphicx}
\usepackage{caption} %

\usepackage{tikz}
\usetikzlibrary{trees}
\usetikzlibrary{calc}
\usetikzlibrary {arrows.meta} 
\usepackage{lipsum}
\usepackage{siunitx}
\usepackage{varwidth}
\usepackage{pifont}
\usepackage{cancel}
\newcommand{\xdownarrow}[1]{%
  {\left\downarrow\vbox to #1{}\right.\kern-\nulldelimiterspace}
}

\usetikzlibrary{shapes.geometric, arrows, shadows}

\tikzstyle{arrow} = [thick,->,>=stealth]
\tikzstyle{arrow1} = [thick,<->,>=stealth]
\tikzstyle{line} = [draw, -latex']
\tikzset{
  double arrow/.style args={#1 colored by #2 and #3}{ thick,
    -latex,  line width=1.1*(#1),#2, %
    postaction={draw,-latex,#3,line width=1.9*(#1/2),
      shorten <=0.7*(#1)/3,shorten >=(#1)/3}, %
  }
}

\advance\textheight\topskip
\numberwithin{equation}{section} \makeatletter
\@addtoreset{equation}{section}

\newcommand{\D}{\text{d}}

\definecolor{ggreen}{rgb}{0.0, 0.6, 0.0}

\begin{document}

\title{Lessons from discrete light-cone quantization for physics at null 
  infinity: Bosons in two dimensions}

\author{Glenn Barnich, Sucheta Majumdar, Simone Speziale, Wen-Di Tan}

\def\mytitle{Lessons from discrete light-cone quantization for physics at null 
  infinity: Bosons in two dimensions}

\pagestyle{myheadings} \markboth{\textsc{\small G.~Barnich, S. Majumdar,
    S.~Speziale, W. Tan}}
{\textsc{\small DLCQ \& flat holography I}}

\addtolength{\headsep}{4pt}

\begin{centering}

  \vspace{1cm}

  \textbf{\Large{\mytitle}}

  \vspace{.5cm}

    \vspace{1.5cm}

    {\large Glenn Barnich$^a$,
      Sucheta Majumdar$^b$, Simone
      Speziale$^c$, Wen-Di Tan$^a$}

\vspace{1cm}

\begin{minipage}{.9\textwidth}\small \it \begin{center}
    $^a$Physique Th\'eorique et Math\'ematique \\ Universit\'e libre de Bruxelles
  and International Solvay Institutes\\ Campus Plaine C.P. 231, B-1050
  Bruxelles, Belgium \\
  E-mail:
  \href{mailto:Glenn.Barnich@ulb.be}{Glenn.Barnich@ulb.be},
  \href{mailto:wendi.tan@ulb.be}{wendi.tan@ulb.be}
\end{center}
\end{minipage}

\vspace{.5cm}

\begin{minipage}{.9\textwidth}\small \it \begin{center}
    $^b$ ENS de Lyon, CNRS, Laboratoire de Physique, F-69342 Lyon, France \\
    E-mail: \href{mailto:sucheta.majumdar@ulb.be}{sucheta.majumdar@ens-lyon.fr}
  \end{center}
\end{minipage}

\vspace{.5cm}

\begin{minipage}{.9\textwidth}\small \it \begin{center}
    $^c$ Aix Marseille Univ., Univ. de Toulon, CNRS, CPT, UMR 7332\\ 13288 Marseille, France \\
    E-mail: \href{mailto:simone.speziale@cpt.univ-mrs.fr}{simone.speziale@cpt.univ-mrs.fr}
      \end{center}
\end{minipage}

\end{centering}

\vspace{.5cm}

\begin{center}
  \begin{minipage}{.9\textwidth} \textsc{Abstract}. Motivated by issues in the
    context of asymptotically flat spacetimes at null infinity, we discuss in
    the simplest example of a free massless scalar field in two dimensions
    several subtleties that arise when setting up the canonical formulation on a
    single or on two intersecting null hyperplanes with a special emphasis on
    the infinite-dimensional global and conformal symmetries and their canonical
    generators, the free data, a consistent treatment of zero modes, matching
    conditions, and implications for quantization of massless versus massive
    fields.
 \end{minipage}
\end{center}

\vfill
\thispagestyle{empty}

\newpage

{\small 

  \tableofcontents

  }

\vfill
\newpage

\section{Introduction}
\label{sec:introduction}

When studying asymptotically flat spacetimes at future null infinity, outgoing
radiation is readily accessible whereas incoming radiation is somewhat hidden.
In the Bondi-Sachs approach
\cite{Bondi:1960jsa,Bondi:1962px,Sachs1962a,Sachs1962,Madler:2016xju}, it is
contained in the sub-leading components of the angular part of the metric
$g_{AB}$, while in the Newman-Penrose framework
\cite{Newman:1961qr,Newman:1962cia,Penrose1963,Penrose:1965am,newman:1980xx%
  ,newman_spin-coefficient_2009}, it is encoded in the component
$\Psi_0=O(r^{-5})$ of the Weyl tensor (see e.g.~\cite*{Chandrasekhar:1985kt},
pages 520-523 for a discussion of incoming and outgoing waves in this context).
This asymmetry is due to the fact that the standard coordinate systems used in
both of these approaches involve a null retarded time coordinate $u$ and a
radial coordinate $r$ that controls the asymptotics, the area-radius in the
former and the affine parameter for null generators in the latter. In order to
achieve an understanding of scattering in this framework, the analysis is
repeated by using a null advanced time coordinate $v$ and exchanging the role of
incoming and outgoing radiation, with important antipodal matching conditions
\cite{Strominger:2013jfa}. Rather than discovering the effects of incoming,
respectively outgoing, radiation layer by layer in an asymptotic expansion, it
seems more efficient to cut through the layers and use double-null light-cone
coordinates to set up the problem (see e.g.~\cite{Wald:1984rg}, section 11.1 for
an introduction).

Since asymptotically flat general relativity is to a large extent controlled by
the linearized theory, we will at this stage not attempt directly to identify
incoming and outgoing gravitons in full general relativity (see
e.g.~\cite{Penrose:1976jq}), but instead point out subtleties that arise when
setting up the canonical formalism for free massless fields on a null
hypersurface. Whereas in the instant form of dynamics, first class constraints
generate gauge symmetries under standard assumptions, this is not necessarily
the case in the front form: the first class constraints associated to the
definition of the canonical momenta generate infinite-dimensional global
symmetries of the type that have attracted considerable attention in recent
investigations. This is ultimately due to the fact that the initial data that
one fixes on a null hyperplane do not determine the wave that moves along the
direction of this hyperplane (see e.g.~\cite{Dirac:1949cp}, section 6, paragraph
2).

In the present note, we explicitly study these subtleties in all detail in the
well-understood and completely tractable example of a free massless scalar in
$1+1$ spacetime dimensions. As emphasized for instance in
\cite{Pauli:1985pv,Pauli:1985ps,Mccartor1988,Heinzl:1991vd,Heinzl1994}, see also
\cite{Brodsky1998,Heinzl:2000ht} for general reviews, the $1+1$ dimensional case
serves not only as a proof of concept, but the results are also an important
ingredient when considering spin 0,1 and 2 fields on a flat four dimensional
Minkowski background, where the intersection of null hyperplanes is
two-dimensional Euclidean space rather than a point. Our aim is thus not to
learn something new about the free massless boson but rather about the canonical
formalism on null hypersurfaces.

We start by studying the canonical formulation on a single light front
(hereafter simply `front'), i.e., a null line where initial data is provided.
Whereas in the instant form of dynamics, a free massless scalar field is
equivalent to a left and a right moving chiral boson, together with a
(space-independent) zero mode (see e.g.
\cite{Floreanini:1987as,Bernstein:1988zd,Sonnenschein:1988ug,Salomonson:1988mk,%
  Stone:1989vg,Henneaux1989}), in the front form it is equivalent to a single
chiral boson, together with a zero mode sector. As shown in
\cite{Alexandrov:2014rta}, this sector is associated with a first class
constraint that appears in connection with the well-known second class
constraint of the light-front analysis. This sector contains a Lagrange
multiplier with an arbitrary time dependence that encodes the data on the other
front, in-line with a characteristic initial value problem. Even though this
sector has the same form than that of a canonical pair that is pure gauge, the
main point is that it has to be interpreted differently because it contains the
wave moving along the front. What is missing in this approach is a Poisson
bracket structure in this sector.

Once it becomes clear that half of the physical degrees of freedom of the
problem, e.g., the left movers if choosing $x^+$ as evolution time, are encoded
in a zero mode, it is convenient to set up the problem on a finite $x^-$
interval in order to cleanly separate the zero from the other modes, before
taking the large interval limit if needed. This procedure of putting the system
in a null box requires suitable boundary conditions. The case of periodic or
anti-periodic ones is referred to as discrete light cone quantization
(DLCQ)\footnote{The formulation is actually based on a light-plane, hence a
  front, and not on a light cone.} in the literature
\cite{Maskawa:1975ky,Pauli:1985pv,Pauli:1985ps}. Such boundary conditions might
be considered as a bounded version of ``Christodoulou-Klainerman'' fall-off
conditions in retarded time $u$ on the news tensor that have been considered in
the context of gravitational scattering theory.

If one studies the system on the standard time-like cylinder, the treatment
becomes more involved because the particle zero mode of the instant form does
not satisfy the DLCQ periodicity conditions. Due care is needed in order to
recover the physics of this mode from the light-front analysis. This is where
improvement terms of Regge-Teitelboim / Gibbons-Hawking
\cite{Regge:1974zd,Gibbons:1976ue} type become relevant in order to have
differentiable symmetry generators and a well-defined action principle. 

We then proceed to show how a formulation on two intersecting fronts (see
e.g.~\cite{dInverno:1980kaa,Nagarajan1985,Mccartor1988,Hayward1993}) allows one
to treat both types of waves in a symmetrical way, and provides the correct
canonical structure for all variables. A convenient description that is
extensively used in the literature on light-cone quantization of free fields is
to use the Peierls bracket, or its quantum version, the Pauli-Jordan commutation
function. It allows one to directly work on the level of the general solution to
the equations of motion, which is explicitly known. This is not a necessity
however. If one chooses to work with standard equal time Poisson / Dirac
brackets on the two fronts special care has to be devoted to suitable matching
conditions so that the two particle zero modes of both fronts combine, in the
case of the time-like cylinder, into the usual single particle zero mode in
instant form.

In order to get correct interpretations in both formulations and ensure
equivalence with quantization in instant form, our guiding principles are (i)
first order action principles that are related to the second order Lagrangian
ones through elimination of auxiliary fields (see e.g.~\cite{Henneaux:1992ig}),
and (ii) an application of Stokes' theorem in the case of a spacetime volume
bounded by an initial surface used in instant form and intersecting fronts
\cite{Neville:1971zk}.

As a consistency check, we provide in double front form, the mode expansion, the
quantization and the computation of the well-known partition function in the
case of periodic boundary conditions on a spatial interval. It is interesting to
reproduce this specific observable in this approach because there is a tension:
while solutions and symmetries simplify in light-cone coordinates, the boundary
conditions are not well adapted, and are natural in instant form. Only a
detailed understanding of all sectors, i.e., left and right movers, together
with particle zero modes and their matching conditions, yields correct results.

For comparison, we briefly recall in an appendix dedicated to the standard
instant form of dynamics, how a free massless Klein-Gordon field in two
dimensions splits off-shell, on the level of action principles, into a left and
a right moving chiral bosons, together with a particle zero mode. We also
provide the well-known partition function that we are reproducing in the main
text through quantization on intersecting double fronts, as well as the
non-trivial mixing through conformal transformations of the left and right
movers with the zero mode sector.

In a section devoted to the massive case, we quickly review the well-known
result that, as a direct consequence of Dirac's algorithm, there is no zero
mode, and that for periodic boundary conditions on the front, it is enough to
provide data on a single front, in line with the analysis of
\cite{Mccartor1988,Heinzl1994}, while in the final section, we summarize the
main lessons that can be drawn from our analysis.

Recent work that deals with infinite-dimensional symmetries or the symplectic
structure in the light-cone approach on null hypersurfaces includes for instance
\cite{Chandrasekaran:2018aop,Ananth:2020ngt,Ananth:2020ojp,Ashtekar:2021kqj,%
  Krishnan2022,Krishnan2023,Majumdar2023,%
  Gonzalez2023,Ciambelli:2023mir,Odak:2023pga,Adami:2023wbe,Chandrasekaran:2023vzb}.

\section{Light-front Lagrangian analysis}
\label{sec:massl-scal-field}

\subsection{Dynamics}
\label{sec:dynamics}

In light-cone coordinates,
\begin{equation}
  \label{eq:1}
  x^\pm=\frac{x^0\pm x^1}{\sqrt 2},
\end{equation}
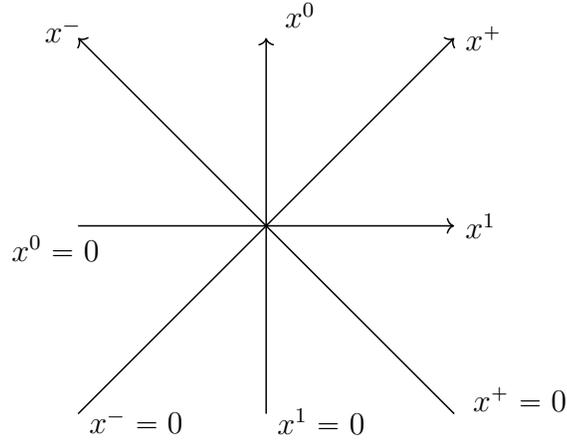
\begin{figure}[!htb]
  \centering
  \begin{tikzpicture}[dot/.style={circle,inner sep=1pt,fill,label={#1},name=#1},
    extended line/.style={shorten >=-#1,shorten <=-#1},
    extended line/.default=1cm]
    \draw[<-, line width = 0.2mm] (0,2.5) -- (0,-2.5) node[right, yshift=-0.1cm] {$x^1=0$} node[right, yshift= 5.3cm, xshift= 1mm]{$x^0$};
    \draw[<-, line width = 0.2mm] (2.5,2.5) -- (-2.5,-2.5) node[right, yshift= -0.1cm] {$x^-= 0$} node[right, yshift= 5cm, xshift=5cm]{$x^+$}; 
    \draw[<-, line width = 0.2mm] (-2.5,2.5) -- (2.5,-2.5) node[right, yshift=0.2cm, xshift=0.1cm] {$x^+=0$} node[yshift= 5.1cm, xshift= -5.2cm]{$x^-$};
    \draw[->, line width = 0.2mm] (-2.5,0) -- (2.5,0) node[right] {$x^1$} node[xshift = -5.3cm, yshift= -.3cm]{$x^0=0$};
  \end{tikzpicture}
  \caption{Coordinate axes} \label{fig:grid}
\end{figure}
the two dimensional Minkowski metric and the action
\begin{equation}
  \label{eq:3}
  ds^2=(dx^0)^2-(dx^1)^2,\quad S[\phi]=\frac 12 \int dx^0dx^1\,
  \partial_\mu\phi\partial^\mu\phi,
\end{equation}
become
  \begin{equation}
    \label{eq:2}
    ds^2=2dx^+dx^-,\quad S=\int dx^+dx^-\, \mathcal{L},\quad \mathcal{L}
    =\partial_+\phi\partial_-\phi.
  \end{equation}
  The Euler-Lagrange equations of motion are
\begin{equation}
  \label{eq:36}
  \partial_+\partial_-\phi=0,
\end{equation}
with general solution given by left and right moving waves,
\begin{equation}
  \label{eq:4}
  \phi=\phi^S_+(x^+)+\phi^S_{-}(x^-),
\end{equation}
that propagate in the $x^-$ and the $x^+$ direction, respectively. In terms of initial
conditions on the light fronts, i.e., the null lines $x^\pm=c^\pm$,
\begin{equation}
  \label{eq:34}
  \phi^S_+(x^+)=\phi(x^+,c^-),\quad\phi^S_{-}(x^-)=\phi(c^+,x^-),
\end{equation}
together with the matching condition
\begin{equation}
  \label{eq:35}
  \phi^S_{+}(c^+)=\phi^S_{-}(c^-),
\end{equation}
at the intersection of the light fronts.

When using the freedom to choose the initial value null lines at a given fixed
time, $x^\pm=c^\pm$, one can make them intersect at any point in two dimensional
Minkowski spacetime. Possible choices would be for instance the origin, spatial
infinity, as appropriate for studying asymptotically flat spacetimes at null and
spatial infinity, or future/ past timelike infinity, where they appear as
discontinuous deformations of a standard Cauchy problem, i.e., where the
initial data is given ``in instant form''.

\begin{figure}[!htb]
\centering
\begin{tikzpicture}[dot/.style={circle,inner sep=1pt,fill,label={#1},name=#1},
  extended line/.style={shorten >=-#1,shorten <=-#1},
  extended line/.default=1cm]
    \draw[-, line width = 0.2mm] (-6.5,1.5) -- (-3.5,1.5) node[yshift= 3mm, midway]{\footnotesize{Origin O}};
  \draw[-, line width = 0.2mm] (-6.5,1.5) -- (-6.5,-1.5);
    \draw[-, line width = 0.2mm] (-6.5,-1.5) -- (-3.5,-1.5) node[midway, yshift= -5mm] {(a)};
      \draw[-, line width = 0.2mm] (-3.5,1.5) -- (-3.5,-1.5);
\draw[-, line width = 0.2mm, green] (-3.5,1.5) -- (-6.5,-1.5);
\draw[-, line width = 0.2mm, magenta] (-6.5,1.5) -- (-3.5,-1.5);
\draw[-, line width = 0.1mm, gray] (-6.5,0) -- (-3.5,0) ;
  \draw[-, line width = 0.2mm] (-1.5,1.5) -- (-1.5,-1.5);
  \draw[-, line width = 0.2mm] (-1.5,1.5) -- (1.5,1.5) node[yshift= 3mm, midway]{\footnotesize{Spatial Infinity $i^0$}};
    \draw[-, line width = 0.2mm] (-1.5,1.5) -- (1.5,1.5);
      \draw[-, line width = 0.2mm] (1.5,1.5) -- (1.5,-1.5);
        \draw[-, line width = 0.2mm] (1.5,-1.5) -- (-1.5,-1.5) node[midway, yshift= -5mm] {(b)};
\draw[-, line width = 0.1mm, gray] (1.5,0) -- (-1.5,0);
\draw[-, line width = 0.2mm, magenta] (0,1.5) -- (1.5,0);
\draw[-, line width = 0.2mm, green] (0,-1.5) -- (1.5,0);
  \draw[-, line width = 0.2mm] (6.5,1.5) -- (3.5,1.5) node[yshift= 3mm, midway]{\footnotesize{Timelike Infinity $i^+,i^-$}};
  \draw[-, line width = 0.2mm] (6.5,1.5) -- (6.5,-1.5);
    \draw[-, line width = 0.2mm] (6.5,-1.5) -- (3.5,-1.5) node[midway, yshift= -5mm] {(c)};
      \draw[-, line width = 0.2mm] (3.5,1.5) -- (3.5,-1.5);
\draw[-, line width = 0.2mm, green] (5,1.5) -- (3.5,0);
\draw[-, line width = 0.2mm, magenta] (5,1.5) -- (6.5,0);
\draw[-, line width = 0.1mm, yellow] (5,-1.5) -- (3.5,0);
\draw[-, line width = 0.1mm, orange] (5,-1.5) -- (6.5,0);
\draw[-, line width = 0.1mm, gray] (3.5,0) -- (6.5,0);
\end{tikzpicture}
\caption{Intersecting initial null planes}  \label{fig:intersections}
\end{figure}
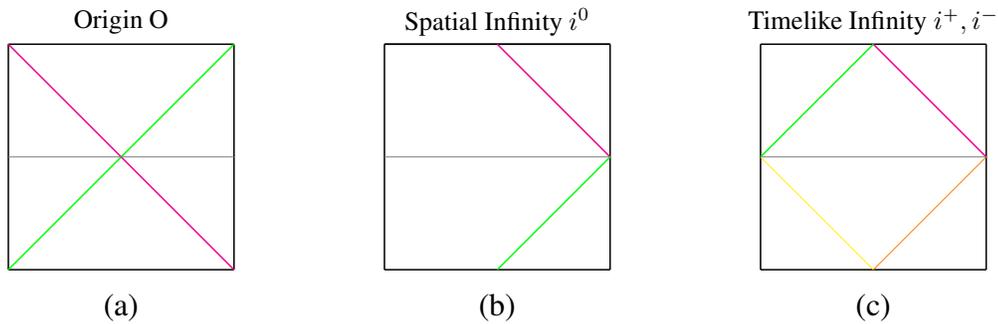

\subsection{Symmetries and Noether currents}
\label{sec:symmetries-1}

Our convention for global symmetries and Noether currents are as follows. If
$\delta_Q\phi^i=Q^i$ is a symmetry of the action, $\delta_Q
\mathcal{L}=\partial_\mu k^\mu_Q$ for some $k^\mu_Q$, the Noether current for a
Lagrangian that depends at most on first order derivatives of the fields is
$j^\mu_Q=\frac{\partial \mathcal{L}}{\partial \partial_\mu \phi^i}Q^i-k^\mu_Q$
and satisfies
\begin{equation}
  Q^i\frac{\delta \mathcal{L}}{\delta\phi^i}+\partial_\mu j^\mu_Q=0.
\end{equation}

The correspondence between global symmetries and Noether currents is unique only
when defining equivalence classes (see e.g.~\cite{Barnich:2000zw,Barnich:2001jy}
for details). In the current context where there are no non-trivial gauge
symmetries, global symmetries of the form
\begin{equation}
  \label{eq:trivial}
  Q^i_T=M^{[ij]}\frac{\delta \mathcal{L}}{\delta\phi^j}+M^{i,j\mu}\partial_\mu\frac{\delta \mathcal{L}}{\delta\phi^j}
  +\partial_\mu\big[M^{j,i\mu}\frac{\delta \mathcal{L}}{\delta\phi^j}\big]-\partial_\mu[M^{[i\mu,j\nu]}\partial_\nu\frac{\delta \mathcal{L}}{\delta\phi^j}\big],
\end{equation}
where $M^{[ij]}=-M^{[ji]}$, $M^{[i\mu,j\nu]}=-M^{[j\nu,i\mu]}$ with associated
Noether currents
\begin{equation}
  \label{eq:trivialNoether}
  j^\mu_{Q_T}=-\frac{\delta \mathcal{L}}{\delta\phi^i}M^{j,i\mu}\frac{\delta \mathcal{L}}{\delta\phi^j}
  +\frac{\delta \mathcal{L}}{\delta\phi^i}M^{[i\mu,j\nu]}\partial_\nu \frac{\delta \mathcal{L}}{\delta\phi^j},
\end{equation}
are trivial. More generally, on the level of currents, weakly vanishing currents
and currents that are divergences of superpotentials, $j^\mu=\partial_\nu k^{[\mu\nu]}$, are
trivial. As we will see below, additional considerations may require specific
``improved'' representatives.

If $\epsilon^+=\epsilon^+(x^+)$, $\epsilon^-=\epsilon^-(x^-)$, the chiral shift transformations
\begin{equation}
  \label{eq:26}
  \delta_{\epsilon^\pm}\phi=\epsilon^{\pm},
\end{equation}
map solutions to solutions. Furthermore, they are global symmetries of the
action since
\begin{equation}
    \label{eq:22}
    \delta_{\epsilon^+} \mathcal{L}=\partial_-(\phi\partial_+\epsilon^+),\quad \delta_{\epsilon^-}\mathcal{L} =\partial_+(\phi\partial_-\epsilon^-),
\end{equation}
and the associated Noether currents are
\begin{equation}
    \label{eq:20}
    j^+_{L,\epsilon^+}=\partial_-\phi\epsilon^+,\ j^-_{L,\epsilon^+}=\partial_+\phi\epsilon^+-\phi\partial_+\epsilon^+,\
    j^+_{L,\epsilon^-}=\partial_-\phi\epsilon^--\phi\partial_-\epsilon^-, \ j^-_{L,\epsilon^-} =\partial_+\phi\epsilon^-.
  \end{equation}
In particular, the chiral shifts contain the constant shift $\delta_c\phi=c$ with Noether
current $j^\mu_c=\partial^\mu \phi c$. 

Furthermore, the Noether currents associated to conformal symmetries,
\begin{equation}
  \label{eq:27}
  \delta_\xi\phi=\xi^\rho\partial_\rho\phi,\quad \xi^+=\xi^+(x^+),\ \xi^-=\xi^-(x^-),
\end{equation}
are given in terms of the energy momentum tensor
\begin{equation}
  \label{eq:28}
  T_{\pm\pm}=(\partial_\pm\phi)^2,\quad T_{\pm\mp}=0,
\end{equation}
as
\begin{equation}
    \label{eq:29}
    j^+_{L,\xi}=\xi^- T_{--},\quad j^-_{L,\xi}=\xi^+ T_{++}.
\end{equation}

\section{Single front Hamiltonian analysis}
\label{sec:light-cone-hamilt}

\subsection{Dynamics}
\label{sec:dynamics-1}

We start by following the discussion in \cite{Heinzl:2000ht,Alexandrov:2014rta}.
We will mark with a $^+$ all quantities to keep track of our choice of evolution
time. This is not necessary in this section and makes the notation cumbersome,
but it will pay off in the next sections when we will consider both fronts at
once. When taking $x^+$ as the direction of evolution, the definition of the
canonical momentum $\pi^+$ on the $x^+=c^+$ front leads to the primary
constraints
\begin{equation}
    \label{eq:5}
    g^+=\pi^+-\partial_-\phi\approx 0,
\end{equation}
which hold at each point $x^-$. The canonical Hamiltonian,
\begin{equation}
  \label{eq:38}
  G^+[\lambda^+]=\int dx^- g^+\lambda^+,
\end{equation}
vanishes on the primary constraint surface.
The first order constrained action that is equivalent to the
Lagrangian one by elimination of the auxiliary fields given by the canonical
momentum $\pi^+(x^+,x^-)$ and the Lagrange multiplier $\lambda^+(x^+,x^-)$ is
then
\begin{equation}
      \label{eq:7}
      S_H[\phi,\pi^+,\lambda^+]=\int dx^+\int dx^-\ \mathcal{L}_H^+,\quad \mathcal{L}_H^+=\pi^+ \partial_+\phi
      -\lambda^+(\pi^+-\partial_-\phi).
\end{equation}

By construction, the solution of the associated Euler-Lagrange
equations of motion
\begin{equation}
    \label{eq:16}
    \pi^+=\partial_-\phi,\quad \lambda^+=\partial_+\phi,\quad \partial_+\pi^++\partial_-\lambda^+=0,
\end{equation}
give not only \eqref{eq:4}, but fix also the auxiliary fields in terms of this
solution,
\begin{equation}
    \label{eq:17}
    \pi^+=\partial_-\phi^{S}_-,\quad \lambda^+=\partial_+\phi^{S}_+.
\end{equation}

Let us now see how these equations appear in terms of the canonical structure.
The non-vanishing equal $x^+$ time Poisson bracket associated to the first order
action \eqref{eq:7} is
\begin{equation}
  \label{eq:9}
  \{\phi(x^+,x^-),\pi^+(x^+,y^-)\}_+=\delta(x^-,y^-).
\end{equation}
The Poisson brackets of the constraints smeared by Lagrange
multipliers,
are given by
\begin{equation}
    \label{eq:11}
    \{G^+[\lambda^{+1}], G^+[\lambda^{+2}]\}_+=\int
    dx^-(\lambda^{+2}\partial_-\lambda^{+1}
    -\lambda^{+1}\partial_-\lambda^{+2}).
\end{equation}
When following Dirac's algorithm, there are no secondary constraints. Instead
there are restrictions on the Lagrange multipliers from the requirement that
\begin{equation}
    \label{eq:8}
    \{g^+(x^+,x^-),\int dy^- \lambda^+(x^+,y^-)
    g^+(x^+,y^-)\}_+=-2\partial_-\lambda^+\approx 0
    \Longrightarrow \partial_-\lambda^+=0,
\end{equation}
so that $\lambda^+=\bar \lambda^+_+(x^+)$. This implies that the $x^-$ zero modes of the local
constraint $g^+$,
\begin{equation}
  \label{eq:10}
  \bar g^+_+=\int dy^- g^+,
\end{equation}
is first class, and so is
\begin{equation}
  \label{eq:12}
  G^+[\bar \lambda^+]=\bar \lambda^+_+\bar g^+_+.
\end{equation}
By construction, the Hamiltonian equations of motion that are equivalent to the
Lagrangian ones are the constraints \eqref{eq:5} together
with the evolution equations
\begin{equation}
  \label{eq:32}
  \partial_+\phi=\{\phi,G^+[\bar \lambda^+]\}_+=\bar \lambda^+_+,\quad \partial_+\pi^+=\{\pi^+,G^+[\bar \lambda^+]\}_+=0.
\end{equation}

In terms of $\bar \lambda^+_+(x^+)$, the general solution of the evolution equation in
$x^+$ is
\begin{equation}
  \label{eq:72}
  \phi(x^+,x^-)=\int^{x^+}_{c^+}dy^+ \bar\lambda^+_+ (y^+)+\phi(c^+,x^-).
\end{equation}
When using the last of \eqref{eq:32}, the constraints \eqref{eq:5} are solved in
terms of $\pi^+(x^-)$ as
\begin{equation}
    \label{eq:73}
    \phi(x^+,x^-)=\int^{x^-}_{c^-}dy^- \pi^+(y^-)+\phi(x^+,c^-).
\end{equation}
Matching at either $c^+$ or $c^-$ then gives
\begin{equation}
    \label{eq:74}
    \phi(x^+,x^-)=\int^{x^+}_{c^+}dy^+ \bar \lambda^+_+(y^+)+\int^{x^-}_{c^-}dy^- \pi^+(y^-)+\phi(c^+,c^-).
\end{equation}
The general solution is thus determined by the free data
$\bar \lambda^+_+(x^+)$ at $x^-=c^-$, $\pi^+(x^-)$ at $x^+=c^+$ and a constant at the
intersection.

Two lessons can be drawn from this analysis:

\begin{itemize}
\item {\em The Dirac algorithm on a null hyperplane provides the correct free data
  for the characteristic initial value problem.}

\item {\em Contrary to what happens in the instant form where the Lagrange multipliers for
  first class constraints are fixed on-shell in terms of pure gauge degrees of
  freedom, they now encode information on the physical degree of freedom that
  cannot be seen on the front. These constraints have thus to be treated differently
  than first class constraints in instant form.}
\end{itemize}

\subsection{Topology and boundary conditions}
\label{sec:topol-bound-cond}

The previous considerations have been local, independent of the choice of
topology and of boundary conditions. Note that in \eqref{eq:11},
\eqref{eq:8}, one assumes that one may integrate by parts without boundary
terms. The conclusion that follows from this assumption, namely
$\partial_-\lambda^+=0$, can however also be reached directly from the
Hamiltonian equations of motion, \eqref{eq:16}, without any restrictions.

If one wants to discuss Noether's first theorem, not on a local level as in section
\ref{sec:symmetries-1}, but in terms of canonical generators, topology and
boundary conditions do become important. The same applies if one
requires a well-defined variational principle, i.e., if one insists
that variation of the action produces the Euler-Lagrange equations
of motion, without any boundary terms.

\begin{figure}
\centering
\begin{tikzpicture}[dot/.style={circle,inner sep=1pt,fill,label={#1},name=#1},
  extended line/.style={shorten >=-#1,shorten <=-#1},
  extended line/.default=1cm]
  \draw[<-, line width = 0.1mm] (0,4.1) -- (0,-4.1) node[right, yshift= 8cm]{$x^0$} node[yshift=-0.4cm] {$x^1=0$};
\draw[<-, line width = 0.1mm] (4,4) -- (-4,-4) node[right, yshift= -0.1cm] {$x^-= 0$} node[right, yshift= 8cm, xshift=8cm]{$x^+$}; 
\draw[<-, line width = 0.1mm] (-4,4) -- (4,-4) node[right, yshift=0.5cm, xshift=-.2cm] {$x^+=0$} node[yshift= 8.1cm, xshift= -7.5cm]{$x^-$};
\draw[->, line width = 0.2mm] (-4,0) -- (4,0) node[right] {$x^1$} node[xshift = -8.1cm, yshift= -.3cm]{$x^0=0$};
\draw[-, line width = 0.2mm] (-1,4) -- (-1,-4);
\draw[-, line width = 0.2mm] (1,4) -- (1,-4); 
\draw[-, line width = 0.2mm] (-2,3) -- (3,-2);
\draw[-, line width = 0.2mm] (-2,-3) -- (3,2);
\draw[-, line width = 0.2mm] (-3,-2) -- (2,3);
\draw[-, line width = 0.2mm] (-3,2) -- (2,-3);
\draw[-, line width = 0.5mm, blue] (-1,0.0) -- (0,1) node[midway, yshift=5mm, xshift=-0.7mm]{$\frac{L}{\sqrt 2}$};
\draw[-, line width = 0.5mm, red] (-1,0.0) -- (1,0) node[midway, xshift= 1.1mm, yshift = -3.2mm]{$L$} ;
\draw[-, line width = 0.5mm, blue ] (0,1) -- (1,0) node[midway, yshift=5mm, xshift=0.4mm]{$\frac{L}{\sqrt 2}$};
\end{tikzpicture}
\caption{Periodicities} \label{fig:period}
\end{figure}
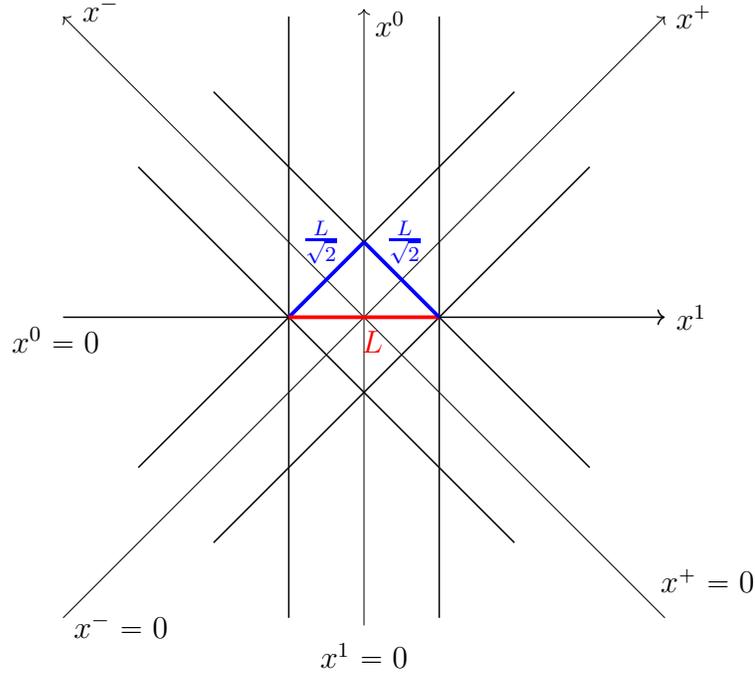

In DLCQ, periodic or anti-periodic boundary conditions are imposed on a finite
interval along a null direction (see e.g.~\cite{Lenz:1991sa,Martinovic2007} for
further discussions on the consistency of such conditions). What we are
interested in here is how suitable modifications of these conditions allow one
to describe all the degrees of freedom of the system on the standard time-like
cylinder, described by $x^0\in \mathbb R$, $x^1\sim x^1+L$. As illustrated by figure
\ref{fig:period}, in light-cone coordinates, the periodic boundary conditions
$x^1\sim x^1+L$ translate into the entangled periodicity conditions
\begin{equation}
  \label{eq:119}
  (x^+,x^-)\sim (x^++L_+,x^--L_-),\quad L_\pm =\frac{L}{\sqrt 2}.
\end{equation}
One may then make a Fourier expansion of the field $\phi$ in the coordinate
$x^1$, and the general solution to the equations of motion can be written as
\begin{equation}
  \label{eq:221}
  \begin{split}
  \phi&=\bar\phi_0(0)+\bar\pi^0_0(0)\frac{x^0}{L}+\phi^R(x^-)+\phi^L(x^+)\\&=
    \bar\phi_0(0)+\bar\pi^0_0(0)\frac{x^++x^-}{\sqrt 2L}+\phi^R(x^-)+\phi^L(x^+),
  \end{split}
\end{equation}
where the left and right movers, $\phi^R(x^-)$ and $\phi^L(x^+)$ do not contain zero modes
(see Appendix \ref{sec:decomposition-cylinder} and \ref{sec:mode-expans-part}
for notation and details) and are separately periodic in their arguments. Here,
the pair $(\bar\phi_0,\bar\pi^0_0)$ are the canonical variables for the
particle sector of the instant form. 

The point is that the general solution is not periodic in $x^-$ at all values of
$x^+$ unless $\bar\pi^0_0(0)=0$. Its derivatives,
$\partial_-\phi=\pi^+,\partial_+\phi=\lambda^+$ on the other hand, are periodic
in $x^-$ at all values of $x^+$. Therefore the description of the particle
sector of the instant form is related to non-periodic boundary conditions in the
front form, and thus to boundary degrees of freedom. The non-periodicity may be
characterized by the quantity
\begin{equation}
  \label{eq:124}
  \Delta_+(x^+) = \phi(x^+,L_-/2)-\phi(x^+,-L_-/2)= \int^{{L_-}/{2}}_{-{L_-}/{2}} dx^- \partial_-\phi.
\end{equation}
Furthermore, a general off-shell field may be written as
\begin{equation}
  \label{eq:24}
  \phi=\Delta_+\frac{x^-}{L_-}+\varphi,
\end{equation}
where $\varphi$ is periodic in $x^-$. 

On-shell, $\Delta_+= \frac 12\bar \pi^0_0(0)$, so that:

{\em If one works in the context of discrete light cone quantization, that is to
  say with a light-like cylinder described by $x^-\sim x^-+L_-$ at all values of $x^+$,
  the solution space will be restricted to the sector where the momentum of the
  particle zero mode associated to the time-like cylinder vanishes.}

In the following, we thus assume that $\phi$ satisfies the boundary conditions
specified in \eqref{eq:24}, while $\pi^+,\lambda^+$ are periodic in $x^-$. Even
though in the general case, the non-periodicity is parametrized by
$\Delta_+(x^+)$, we will furthermore assume at this stage that $\Delta_+$ is
constant, since this is what it is on-shell. This guarantees that one does not
lose any solutions associated to the time-like cylinder. The case where
$\Delta_+=0$ is referred to as the periodic case.

Note that the Hamiltonian $G^+[\lambda^+]$ is not a differentiable generator in the
non-periodic case in the sense of \cite{Regge:1974zd} and as consequence, the
first order action principle is not well-defined. Indeed, 
\begin{equation}
  \label{eq:216}
  \delta G^+[\lambda^+]=\int_{-L_-/2}^{L_-/2}dx^-[\delta\lambda^+(\pi^+-\partial_-\phi)+\delta\pi^+\lambda^++\partial_-\lambda^+\delta\phi]-\oint_{\partial \mathcal V}\lambda^+\delta\phi,
\end{equation}
where $\mathcal V=[-L_-/2,L_-/2]$. We will show below how this issue is addressed.

\subsection{Zero mode and chiral boson sectors}
\label{sec:dirac-brackets}

The $x^-$ zero mode of the Lagrange multiplier is defined as
\begin{equation}
  \label{eq:13}
  \bar \lambda_+^+(x^+)=\frac{1}{L_-}\int^{{L_-}/{2}}_{-{L_-}/{2}} dx^- \lambda^+ (x^+,x^-),
\end{equation}
so that
\begin{equation}
  \label{eq:39}
  \lambda^+= \bar \lambda^+_+ + \tilde \lambda_+^+,\quad \int^{{L_-}/{2}}_{-{L_-}/{2}} dx^- \tilde\lambda_+^+=0.
\end{equation}
The canonical variables may be decomposed in the same way as the Lagrange multipliers,
\begin{equation}
  \label{eq:41}
  \begin{split}
    & \phi(x^+,x^-)=\bar \phi_+(x^+)+\tilde\phi_+(x^+,x^-),\quad \pi^+(x^+,x^-)=\frac{1}{L_-}\bar
      \pi^+_+(x^+)+
      \tilde \pi^+_+(x^+,x^-), \\
    & \bar \phi_+=\frac{1}{L_-}\int^{{L_-}/{2}}_{-{L_-}/{2}} dx^- \phi,\
      \bar \pi^+_+=\int^{{L_-}/{2}}_{-{L_-}/{2}} dx^- \pi^+,\  \int^{{L_-}/{2}}_{-{L_-}/{2}} dx^-\tilde \phi_+=0
      =\int^{{L_-}/{2}}_{-{L_-}/{2}} dx^-\, \tilde \pi^+_+.
  \end{split}
\end{equation}
Note that the non-periodicity goes into the field without the $x^-$ zero mode, 
\begin{equation}
  \label{eq:25}
  \tilde\phi_+=\Delta_+\frac{x^-}{L_-}+\tilde\varphi_+. 
\end{equation}

The Hamiltonian equations of motion \eqref{eq:16} split as
\begin{equation}
  \label{eq:218}
  \begin{split}
    & \bar\pi^+_+=\Delta_+,\quad\tilde\pi^+_+=\partial_-\tilde\varphi_+,\quad\bar\lambda^+_+=\partial_+\bar\phi_+,\quad\tilde\lambda^+_+=
      \partial_+\tilde\varphi_+,\\
    & \partial_+\bar \pi^+_+=0,\quad\partial_+\tilde\pi^+_++\partial_-\tilde\lambda^+_+=0
  \end{split}
\end{equation}
with general solution given by
\begin{equation}
  \label{eq:219}
  \begin{split}
    & \bar\pi^+_+=\bar\pi^+_+(0)=\Delta_+,\quad\tilde \phi_+=\phi^R(x^-),\quad\tilde\lambda^+_+=0,\\
    & \bar\phi_+
      =\bar\phi_+(0)+\int^{x^+}_0dy^+\bar\lambda^+_+(y^+).
  \end{split}
\end{equation}
Decomposing in turn the $x^-$ zero mode into a constant part and a fluctuating
field,
$\bar\lambda^+_+=\frac{1}{L_+}\int^{L_+/2}_{-L_+/2}dy^+\bar\lambda^+_+(y^+)+\tilde{\bar\lambda}^+_+$,
one finds
\begin{equation}
  \label{eq:220}
  \bar\phi_+ =\bar{\bar\phi}_++\int^{L_+/2}_{-L_+/2}dy^+\bar\lambda^+_+(y^+)\frac{x^+}{L_+}+\partial_+^{-1}\tilde{\bar\lambda}^+_+,
\end{equation}
where $\bar{\bar\phi}_+$ is a constant and $\partial_+^{-1}\tilde{\bar\lambda}^+_+$ is the primitive
of $\tilde{\bar\lambda}^+_+$ that does not contain a constant,
\begin{equation}
  \label{eq:222}
  \partial_+^{-1}\tilde{\bar\lambda}^+_+(x^+)=\int^{x^+}_0dy^+\tilde{\bar\lambda}^+_+(y^+)
  -\frac{1}{L_+}\int^{L_+/2}_{-L_+/2}dy^+\int^{y^+}_0dy^+\tilde{\bar\lambda}^+_+(z^+)=\phi^L(x^+).
\end{equation}
Finally, in order for the solution to have the correct entangled periodicity,
one needs the ``matching'' condition
\begin{equation}
  \label{eq:223}
  \Delta_+=\int^{L_+/2}_{-L_+/2}dy^+\bar\lambda^+_+(y^+),
\end{equation}
and, provided that the integration constants are identified,
$\Delta_+=\frac{\bar\pi^0_0(0)}{2}$, $\bar{\bar \phi}_+=\bar\phi_0(0)$, the
solution to the Hamiltonian equations of motion agrees with
\eqref{eq:221}.

\subsection{Improved action principle}
\label{sec:impr-acti-princ}

A way to get a well-defined action principle from which equations \eqref{eq:218}
derive is the following. We first note that the constraints \eqref{eq:5} split
into the first class constraint \eqref{eq:10} and the second class constraints
\begin{equation}
  \label{eq:40}
  \tilde g^+_+=g^+-\frac{\bar g^+_+}{L_-},\quad \int^{{L_-}/{2}}_{-{L_-}/{2}} dx^-\,\tilde g^+_+ =0,
\end{equation}
where
\begin{equation}
  \label{eq:42}
  \bar g^+_+=\bar \pi^+_+-\Delta_+,\quad \tilde g^+_+=\tilde\pi^+_+-\partial_-\tilde\varphi_+.
\end{equation}
We then set
\begin{equation}
  \label{eq:224}
  \bar\phi_+=\Delta_+\frac{x^+}{L_+}+\bar\varphi_+,
\end{equation}
with $\bar\varphi_+$ periodic in $x^+$. After substitution of these decompositions
into \eqref{eq:7} the improved first order action is
\begin{equation}
  \label{eq:14}
  \begin{split}
    &  S^{I+}_{H}[\bar\varphi_+,\bar \pi^+_+,\bar \lambda^+_{+},\tilde \varphi_+,\tilde\pi^+_+,\tilde \lambda^+_{+};\Delta_+]
       =\int dx^+ L^+_H,,\quad L^+_H=\bar L^+_H + \int^{{L_-}/{2}}_{-{L_-}/{2}} dx^-\,\tilde{\mathcal L}^+_H,\\
   &  \bar L^+_H  =
    \bar \pi^+_+\partial_+\bar\varphi_+-
    \bar \lambda_+^+\bar g^+_++\bar\pi^+_+\frac{\Delta_+}{L_+}-2\frac{\Delta^2_+}{L_+},\\
   &  \tilde{\mathcal L}^+_H =
    \tilde \pi^+_+\partial_+\tilde\varphi_+ -\tilde \lambda^+_+\tilde g^+_+,
  \end{split}
\end{equation}
where the tilded fields are periodic in $x^-$, while the barred fields are
assumed periodic in $x^+$, and the constant $-2\frac{\Delta^2_+}{L_+}$ has been added for
later convenience. 

In terms of the fields of the improved action principle, the non-vanishing equal
$x^+$-time brackets are 
\begin{equation}
  \label{eq:23}
  \{\bar\varphi_+(x^+),\bar\pi^+_+(x^+)\}_+=1,\quad
  \{\tilde\varphi_+(x^+,x^-),\tilde\pi^+_+(x^+,y^-)\}_+=\delta(x^-,y^-)-\frac{1}{L_-}. 
\end{equation}

This variational principle correctly reproduces
all equations of motion, including the matching condition \eqref{eq:223} which
follows from the $x^+$ zero mode part of the equation of motion for
$\bar\pi^+_+$. If one extracts the constant $x^+$ zero mode of the Lagrange
multiplier for the first class constraint,
$\bar\lambda^+_+(x^+)=\bar{\bar\lambda}^+_++\tilde{\bar\lambda}^+_+(x^+)$, this equation splits into the
matching condition \eqref{eq:223}, which reads $\Delta_+=L_+\bar{\bar\lambda}^+_+$ and
$\partial_+\varphi_+=\tilde{\bar\lambda}^+_+$, which contains the right movers. 

The improved action thus decouples into a zero mode sector and the starting
point Lagrangian action \eqref{eq:2} for $\tilde \varphi_+$ instead of $\phi$.
The latter encodes the dynamics of the right moving chiral bosons. If one
forgets that the coordinates $x^\pm$ are null, and identifies $x^+\to x^0$,
$x^-\to x^1$, the former looks like a simple first class constraint system
consisting of a single degree of freedom that is pure gauge because of the first
class constraint $\bar g^+_+\approx 0$. As emphasized before, this
interpretation for the zero mode sector is not correct however, since it
contains the information on the physical relevant left mover, including part of
the particle zero mode of the instant form in the non-periodic case.

A convenient way to restrict the theory described by $\bar L^+_H$ to just the
degrees of freedom of a particle, without the left movers, is to declare the
constant $\Delta_+$ to be dynamical, $\Delta_+=\Delta_+(x^+)$. In this case, solving the constraint
$\bar g^+_+$ in the action yields that of a free particle,
\begin{equation}
  \label{eq:138}
  S^+_P=\int dx^+ [\bar\pi^+_+\partial_+\varphi_+-\frac{(\bar\pi^+_+)^2}{L_+}]. 
\end{equation}

\subsection{Symmetries and generators}
\label{sec:symmetries-2}

Consider first the periodic case. The transformations generated by the
differentiable first class constraint $G^+[\epsilon^+]$, where
$\epsilon^+=\epsilon^+(x^+)$, are
\begin{equation}
  \label{eq:15}
  \delta_{\epsilon^+}\phi=\{\phi,G^+[\epsilon^+]\}_+=\epsilon^+,\quad \delta_{\epsilon^+}\pi^+=\{\pi^+,G^+[\epsilon^+]\}_+=0.
\end{equation}
Even though the transformations are generated by a first class constraints, it has
been argued in \cite{Alexandrov:2014rta} that these transformations cannot be a
gauge symmetries since they transform solutions to inequivalent solutions. The detailed
explanation is as follows:

{\em The expectation that first class constraints generate transformations that
  do not change the physical state is based on the assumption that initial data
  on a constant time surface are enough to fully determine the solution (see
  e.g.~section 1.2.1 of \cite{Henneaux:1992ig}). There is thus no contradiction
  because, in the massless case, fixing data on a single front is not enough to
  completely determine the solution. }

Note also that the mechanism here is different and not directly related to the
case of large gauge transformations where suitably improved differentiable
generators associated with first class constraints do change the physical state
on account of non trivial boundary conditions
\cite{Regge:1974zd,Benguria:1976in}.

At this stage, we go a step further and show that these shift transformations
are global symmetries, both in the periodic and the non-periodic case. In order to
do so, the transformations are extended to the Lagrange multipliers in such a way
that the variation of the integrand of the first order action \eqref{eq:7} is a
divergence (see e.g.~\cite{Henneaux:1992ig}, chapter 3.2.2). This fixes
\begin{equation}
    \label{eq:18}
    \delta_{\epsilon^+}\lambda^+=\partial_+\epsilon^+,\quad \delta_{\epsilon^+}\mathcal{L}_H^+=\partial_-(\phi\partial_+\epsilon^+),
\end{equation}
with associated Noether currents
\begin{equation}
    \label{eq:21}
    j^+_{\epsilon^+}=\pi^+\epsilon^+,\quad j^-_{\epsilon^+} =\lambda^+\epsilon^+-\phi\partial_+\epsilon^+.
\end{equation}
Note that the Hamiltonian form of the currents $j^\pm_{\epsilon^+}$ agree with
the Lagrangian $j^\pm_{L,\epsilon^+}$ in \eqref{eq:20} when using the last two
Hamiltonian equations of motion in \eqref{eq:16}.

In the periodic case, both $G^{+}[\epsilon^{+}]$ and $Q^{+}_{\epsilon^{+}}$ are
differentiable generators. They generate the same symmetry
transformations as $G^+[\epsilon^+]$, while the associated Noether currents
differ by the divergence of a superpotential,
\begin{equation}
    \label{eq:super}
    g^+ \epsilon^+=j^+_{\epsilon^+}-\partial_-(\phi\epsilon^+),
    \quad j'^-_{\epsilon^+}=j^-_{\epsilon^+}+\partial_+(\phi\epsilon^+)=(\lambda^++\partial_+\phi)\epsilon^+.
\end{equation}

The other shift transformations $\delta_{\epsilon^-}\phi=\epsilon^-$ may again be
extended to the auxiliary fields so that the variation of integrand of the first
order action is a divergence,
\begin{equation}
    \label{eq:30}
    \delta_{\epsilon^-}\pi^+=\partial_-\epsilon^-,\quad
    \delta_{\epsilon^-}\lambda^+=0,\quad
    \delta_{\epsilon^-}\mathcal{L}^+_H=\partial_+(\phi\partial_-\epsilon^-),
\end{equation}
with associated Noether currents,
\begin{equation}
  \label{eq:N2s}
  j^+_{\epsilon^-}=\pi^+\epsilon^--\phi\partial_-\epsilon^-,\quad j^-_{\epsilon^-}=\lambda^+\epsilon^-,
\end{equation}
that agree with their Lagrangian counterparts when using the equations of motion
for the auxiliary fields. On the phase space variables, these symmetries are
canonically generated by the associated charges
$Q^+_{\epsilon^-}=\int^{{L_-}/{2}}_{-{L_-}/{2}} dx^- j^+_{\epsilon^-}$,
which are differentiable generators also in the non-periodic case. One
may also define the equivalent Noether currents
\begin{equation}
  \label{eq:46}
  j'^+_{\epsilon^-}=(\pi^++\partial_-\phi)\epsilon^-=j^+_{\epsilon^-}+\partial_-(\phi\epsilon^-),\ j'^-_{\epsilon^-}=j^-_{\epsilon^-}-\partial_+(\phi\epsilon^-)=(\lambda^+-\partial_+\phi)\epsilon^-.
\end{equation}
The associated charges $Q'^{+}_{\epsilon^-}=\int^{{L_-}/{2}}_{-{L_-}/{2}} dx^- j'^+_{\epsilon^-}$ are only
differentiable in the periodic case.

The extension of the conformal symmetries
$\delta_\xi\phi=\xi^\rho\partial_\rho\phi$ to the auxiliary fields is
\begin{equation}
    \delta_\xi \pi^+=\partial_-(\delta_\xi\phi),\quad\delta_\xi\lambda^+=\partial_+(\delta_\xi\phi),
    \label{eq:ham}
\end{equation}
with Noether currents
\begin{equation}
    j^+_{\xi}=\pi^+\delta_\xi\phi-\xi^+\partial_+\phi\partial_-\phi,\quad j^-_{\xi}=\lambda^+\delta_\xi\phi-\xi^-\partial_-\phi\partial_+\phi.\label{eq:NCc}
\end{equation}
Representatives for observables such as Noether charges that are adapted to the
Hamiltonian formalism should not involve time derivatives of
the canonical variables. This leads to
\begin{equation}
  \label{eq:cangen}
  Q^{\prime +}_{\xi^+}=\int^{{L_-}/{2}}_{-{L_-}/{2}} dx^- j'^+_{\xi^+},\quad
  Q^{\prime +}_{\xi^-}=\int^{{L_-}/{2}}_{-{L_-}/{2}} dx^- j'^+_{\xi^-}.
\end{equation}
where one finds for the left chiral half of the conformal symmetries,
\begin{equation}
    \label{eq:33}
    \begin{split}
      & j'^+_{\xi^+}=\xi^+\lambda^+(\pi^+-\partial_-\phi),\quad j'^-_{\xi^+}=\xi^+\lambda^+\partial_+\phi,\\
      & \delta'_{\xi^+} \phi=\xi^+\lambda^+=\{\phi,Q^{\prime +}_{\xi^+}\}_+,\quad \delta'_{\xi^+}\pi^+=-\xi^+\partial_-\lambda^+=\{\pi^+,Q^{\prime +}_{\xi^+}\}_+,\\
      & \delta'_{\xi^+}\lambda^+ = \partial_+(\xi^+\lambda^+), 
    \end{split}
\end{equation}
and
\begin{equation}
  \label{eq:37}
  \begin{split}
    & j'^+_{\xi^-}=\xi^-\pi^+\partial_-\phi,\quad j'^-_{\xi^-}=\xi^-\pi^+(\lambda^+-\partial_+\phi),\\
    & \delta'_{\xi^-} \phi=\xi^-\partial_-\phi=\{\phi,Q^{\prime +}_{\xi^-}\}_+,\ \delta'_{\xi^-}\pi^+=\partial_-(\xi^-\pi^+)=\{\pi^+,Q^{\prime +}_{\xi^-}\}_+,\\
  & \delta'_{\xi^-}\lambda^+ = \xi^-\partial_-\lambda^+,
  \end{split}
\end{equation}
for the right chiral half. The generators in \eqref{eq:cangen} are differentiable
only in the periodic case. In addition, the right chiral half of the conformal
algebra is canonically realized in terms of Poisson brackets of the associated
Noether charges, whereas the left is not,
\begin{equation}
  \label{eq:155}
  \{Q^{\prime +}_{\xi_1^-},Q^{\prime +}_{\xi_2^-}\}_+=Q'^+_{[\xi_1^-,\xi^-_2]},\quad
  \{Q^{\prime +}_{\xi_1^+},Q^{\prime +}_{\xi_2^+}\}_+=0\neq Q'^+_{[\xi_1^+,\xi^+_2]}.
\end{equation}

In the non-periodic case, differentiable generators can be obtained on the level
of the improved first order action principle \eqref{eq:14} and its variables.
The left chiral shift symmetries with parameter $\epsilon^+(x^+)$, which we
assume to be periodic in $x^+$, act non-trivially only on the zero mode sector as
\begin{equation}
  \label{eq:225}
  \delta_{\epsilon^+}\bar\varphi^+=\epsilon^+,\quad\delta_{\epsilon^+}\bar\pi^+_+=0,\quad 
  \delta_{\epsilon+}\bar\lambda^+_+=\partial_+\epsilon^+.
\end{equation}
On the phase space
variables, these transformations are canonically generated by the differentiable
generator
\begin{equation}
  \label{eq:217}
  Q^{+I}_{\epsilon^+}=\bar g^+_+\epsilon^+.
\end{equation}

For the right chiral shift symmetries, we require the parameter $\epsilon^-(x^-)$ to be
periodic in $x^-$, and decompose into a constant mode and the non-constant ones, 
$\epsilon^-=\bar \epsilon^-_++\tilde\epsilon^-_+$, with
$\bar \epsilon^-_+=\int_{-L_-/2}^{L_-/2} dx^-\epsilon^-/L_{-}$,
$\int_{-L_-/2}^{L_-/2} dx^-\tilde\epsilon^-_+=0$.
They act as
\begin{equation}
  \label{eq:49}
  \delta_{\epsilon^-}\bar \varphi_+=\bar \epsilon^-_+,\quad\delta_{\epsilon^-}\tilde\varphi_+=\tilde \epsilon^-_+,\quad\delta_{\epsilon^-}\tilde\pi^+_+=\partial_-\tilde \epsilon^-_+,\quad
  \delta_{\epsilon^-}\bar\pi^+_+=0=\delta_{\epsilon^-}\bar \lambda_+^+=\delta_{\epsilon^-}\tilde\lambda^+_+.
\end{equation}
On the phase space variables, these transformations are canonically generated by
the differentiable generators,
\begin{equation}
  \label{eq:227}
  Q^{+I}_{\epsilon^-}=\bar\pi^+_+\bar \epsilon^-_++\int^{L_-/2}_{-L_-/2}dx^- (\tilde \pi^+_++\partial_- \tilde \varphi_+)\tilde\epsilon^-_+. 
\end{equation}

For the conformal symmetries, we assume $\xi^+(x^+),\xi^{-}(x^-)$ to be periodic in the
their arguments. For the left chiral half of the conformal symmetries, we find
\begin{equation}
  \label{eq:45}
  \begin{split}
    &  \delta'_{\xi^+}\bar\varphi_+=\xi^+\bar\lambda^+_+,\quad\delta'_{\xi^+}\tilde\varphi_+=\xi^+\tilde\lambda^+_+,\quad\delta'_{\xi^+}\bar\pi^+_+=0,\quad
      \delta'_{\xi^+}\tilde\pi^+_+=-\xi^+\partial_-\tilde\lambda^+_+,\\
  &  \delta'_{\xi^+}\bar\lambda^+_+=\partial_+(\xi^+\bar\lambda^+_+),\quad \delta'_{\xi^+}\tilde \lambda^+_+=\partial_+(\xi^+\tilde \lambda^+_+),
  \end{split}
\end{equation}
with differentiable canonical generators,
\begin{equation}
  \label{eq:52}
  Q^{I+}_{\xi^+}=\xi^+\bar g^+_+\bar\lambda^+_++\xi^+\int^{L_-/2}_{-L_-/2}dx^- \tilde\lambda^+_+(\tilde\pi^+_+-\partial_-\tilde\varphi_+), 
\end{equation}
for the transformations on the phase space variables. They do not provide a
canonical realization of the left chiral half of the conformal algebra.

For the right chiral half, the mixing between sectors is more involved.
One finds
\begin{equation}
  \begin{split}
  \label{eq:226}
  &  \delta'_{\xi^-}\bar\varphi_+=\int^{L_-/2}_{-L_-/2}dx^-\frac{\xi^-}{L_-}(\frac{\Delta_+}{L_-}+\partial_-\tilde\varphi_+),\ \delta'_{\xi^-}\bar\pi^+_+=0,\
    \delta'_{\xi^-}\bar\lambda^+_+=\int^{L_-/2}_{-L_-/2}dx^-\frac{\xi^-}{L_-}\partial_-\tilde\lambda^+_+,\\
  &   \delta'_{\xi^-}\tilde\varphi_+=\xi^-(\frac{\Delta_+}{L_-}+\partial_-\tilde\varphi_+)-\int^{L_-/2}_{-L_-/2}dx^-\frac{\xi^-}{L_-}(\frac{\Delta_+}{L_-}+\partial_-\tilde\varphi_+),\\
  &   \delta'_{\xi^-}\tilde\pi^+_+=\partial_-[\xi^-(\frac{\bar\pi^+_+}{L_-}+\tilde\pi^+_+)],\quad \delta'_{\xi^-}\tilde\lambda^+_+=\xi^-\partial_-\tilde\lambda^+_+-\int^{L_-/2}_{-L_-/2}dx^-\frac{\xi^-}{L_-}\partial_-\tilde\lambda^+_+.
  \end{split}
\end{equation}
The associated Noether charges
\begin{multline}
  \label{eq:51a}
  Q^{+}_{\xi^-}=\int_{-L_-/2}^{L_-/2} dx^-\xi^-\tilde\pi^+_+\partial_-\tilde\varphi_+
  +\Delta_+\int_{-L_-/2}^{L_-/2}
  dx^-\frac{\xi^-}{L_-}\tilde \pi^+_+\\+\bar\pi^+_+\int_{-L_-/2}^{L_-/2}
  dx^-\frac{\xi^-}{L_-}\partial_-\tilde\varphi_++\bar\pi^+_+\frac{\Delta_+}{L_-}\int_{-L_-/2}^{L_-/2}
  dx^-\frac{\xi^-}{L_-}, 
\end{multline}
are the improved differentiable canonical generators of the right chiral half of
the conformal transformations on the canonical variables. They agree on-shell
with the charges computed in instant form in \eqref{eq:163}, when the latter is
restricted to the right chiral half and after performing the change of
integration variables $x^1=\sqrt{2} x^-$. Furthermore, they provide a canonical
realization of the right chiral half of the conformal transformations,
\begin{equation}
  \label{eq:86}
  \{Q^+_{\xi^-_1},Q^+_{\xi^-_2}\}=\delta'_{\xi^-_2}Q^+_{\xi^-_1}=Q^+_{[\xi^-_1,\xi^-_2]}. 
\end{equation}

Note that the right chiral half of the shift and the conformal symmetries are
generated on the canonical variables by charges that vanish on the constraint
surface.
\subsection{Reduced theory and Dirac brackets}
\label{sec:dirac-brackets-1}

Solving the second class constraints in the action, i.e., eliminating the
auxiliary fields $\tilde\lambda_+^+,\tilde\pi^+_+$ gives the equivalent reduced
action principle, 
\begin{equation}
  \label{eq:43}
  \begin{split}
    & S_R^+[\bar\varphi_+,\bar\pi^+_+,\bar\lambda^+_+,\tilde\varphi_{+};\Delta_+]=\int_{-L_+/2}^{L_+/2} dx^+ L^+_R,\\
    & L^+_R=\bar L^+_H +\int_{-L_-/2}^{L_-/2} dx^-\, \tilde {\mathcal L}^+_R,\quad 
      \tilde{\mathcal L}^+_R= \partial_-\tilde\varphi_+\partial_+\tilde\varphi_+.
  \end{split}
\end{equation}

One may follow for instance \cite{Maskawa:1975ky}, \cite{Hanson1976}, section
5.D or \cite{Henneaux1989}, section 2.6, and compute the Dirac brackets
associated with the second class constraints. When using that
\begin{equation}
    \{\tilde g^+_+(x^-),\tilde g^+_+(y^-)\}_+=-2\partial_-^{x}\delta(x^-,y^-),
\end{equation}
and
\begin{equation}
  \int dy^- \partial_-^{x^-}\delta(x^-,y^-)\frac{1}{2}\varepsilon(y^--z^-)=\delta(x^-,z^-),
\end{equation}
where $\varepsilon(x)$ is the sign function, it follows that the non-vanishing
Dirac brackets are given by
\begin{equation}
    \label{Diracbrackets}
    \begin{split}
      & \{\tilde \varphi_+(x^-),\tilde \varphi_+(y^-)\}^*_+=-\frac{1}{4}\varepsilon(x^--y^-)
        +\frac{x^--y^-}{2L_-},\\
      & \{\tilde \varphi_+(x^-),\tilde \pi^+_+(y^-)\}^*_+=\frac 12 [\delta(x^-,y^-)-\frac{1}{L_-}],\\
      &  \{\tilde \pi^+_+(x^-),\tilde \pi^+_+(y^-)\}^*_+=\frac 12 \delta'(x^-,y^-),\\
      &\{\bar \varphi_+,\bar \pi^+_+\}^*_+=1,
    \end{split}
  \end{equation}
where by convention $\delta'(x,y)=\partial_x\delta(x,y)=-\partial_y\delta(x,y)$.
More details can be found in Appendix \ref{sec:position-space}.

When working with Dirac brackets, the second class constraints may be imposed
strongly and only the first class constraint $\bar g^+_+\approx 0$ of the zero mode sector
remains. The Hamiltonian reduces to
\begin{equation}
  \label{eq:62}
  H^{+*}=\bar g^+_+\bar \lambda_+^+.
\end{equation}
It vanishes on the first class constraint surface and generates no time
evolution in the chiral sector, but only in the zero mode sector,
\begin{equation}
  \label{eq:44}
  \partial_+\tilde \varphi_+=\{\tilde\varphi_+,H^{+*}\}^*_+=0, \
  \partial_+\bar \varphi_+=\{\bar \varphi_+,H^{+*}\}^*_+=\bar \lambda_+^+,\
  \partial_+\bar \pi^+_+=\{\bar \pi^+_+,H^{+*}\}^*_+=0,
\end{equation}
so that the initial data $\tilde\varphi_+(x^-)$ are constants of the motion.

The generator $Q^{+I}_{\epsilon^+}$ is unaffected by the pull-back to the second class
constraint surface, while the generator $Q^{+I}_{\epsilon^-}$ becomes 
\begin{equation}
    \label{eq:47}
    Q^{+*}_{\epsilon^-}=\bar \pi^+\bar \epsilon^-_++\int_{-L_-/2}^{L_-/2} dx^-\,2\partial_-\tilde \phi \tilde \epsilon^-_+.
\end{equation}
For the left chiral half of the conformal symmetries, the reduced generator is
\begin{equation}
  \label{eq:50}
  Q^{+*}_{\xi^+}=\bar g^+_+{\bar \lambda_+^+} \xi^+.
\end{equation}
It does no longer induce a transformation on $\tilde\varphi_+$ and does not provide a
canonical realization of the left chiral half of the conformal algebra in the
Dirac bracket. 

The reduced transformations for the right chiral half of the conformal
symmetries are given by 
\begin{equation}
  \label{eq:84}
  \begin{split}
    & \delta^R_{\xi^-}\tilde \varphi_+=\xi^-(\frac{\Delta_+}{L_-}+\partial_-\tilde\varphi_+)
      -\int_{-L_-/2}^{L_-/2} dy^-\frac{\xi^-}{L_-}(\frac{\Delta_+}{L_-}+\partial_-\tilde\varphi_+),\\
     &  \delta^R_{\xi^-}\bar \varphi_+=\int_{-L_-/2}^{L_-/2} dy^-\frac{\xi^-}{L_-}(\frac{\Delta_+}{L_-}+\partial_-\tilde\varphi_+),\quad
      \delta^R_{\xi^-}\bar\pi^+_+=0,\\ & \delta^R_{\xi^-}\bar\lambda_+^+=\int_{-L_-/2}^{L_-/2} dy^-\frac{\xi^-}{L_-}\partial_+\partial_-\tilde\varphi_+,
  \end{split}
\end{equation}
and the reduced Noether charge is
\begin{multline}
  \label{eq:51}
  Q^{+*}_{\xi^-}=\int_{-L_-/2}^{L_-/2} dx^-\xi^-(\partial_-\tilde\varphi_+)^2 
  +(\bar\pi^+_++\Delta_+)\int_{-L_-/2}^{L_-/2}
  dx^-\frac{\xi^-}{L_-}\partial_-\tilde\varphi_+\\+\bar\pi^+_+\frac{\Delta_+}{L_-}\int_{-L_-/2}^{L_-/2}
  dx^-\frac{\xi^-}{L_-}. 
\end{multline}

When using that
\begin{equation}
  \frac{\delta L_R^+}{\delta
    \bar\lambda_+^+}\delta^R_{\xi^-}\bar\lambda_+^+ =\bar g^+_+\int_{-L-/2}^{L_-/2} dx^-  \frac{ \xi^-}{2L_-}
  \frac{\delta \tilde{\mathcal L}^+_R}{\delta\tilde\varphi_+},\label{eq:31}
\end{equation}
the equation for the Noether charge,
\begin{equation}
  \label{eq:158}
  \partial_+Q^{+*}_{\xi^-}+ \frac{\delta
    L_R^+}{\delta\bar\varphi_+}\delta^R_{\xi^-}\bar\varphi_++\frac{\delta L_R^+}{\delta
    \bar\lambda_+^+}\delta^R_{\xi^-}\bar\lambda_+^++\int_{-L-/2}^{L_-/2} dx^-\frac{\delta \tilde{\mathcal L}^+_R}{\delta\tilde\varphi_+}
  \delta^R_{\xi^-}\tilde\varphi_+=0,
\end{equation}
may be re-written as
\begin{equation}
  \label{eq:228}
  \partial_+Q^{+*}_{\xi^-}+ \frac{\delta
    L_R^+}{\delta\bar\varphi_+}\hat \delta^R_{\xi^-}\bar\varphi_++\int_{-L-/2}^{L_-/2} dx^-\frac{\delta \tilde{\mathcal L}^+_R}{\delta\tilde\varphi_+}
  \hat \delta^R_{\xi^-}\tilde\varphi_+=0,
\end{equation}
where the ``improved'' transformations are defined as  
\begin{equation}
  \label{eq:229}
  \begin{split}
   \hat \delta^R_{\xi^-}\bar \varphi_+ & =\delta^R_{\xi^-}\bar\varphi_+,\quad \hat \delta^R_{\xi^-}\bar \pi^+_+=0=\hat \delta^R_{\xi^-}\bar\lambda^+_+,\\
   \hat \delta^R_{\xi^-}\tilde\varphi_+ & =\delta^R_{\xi^-}\tilde\varphi_++\bar g^+_+
                         (\frac{\xi^-}{2L_-}-\int_{-L-/2}^{L_-/2} dy^-\frac{\xi^-}{2L_-^2})\\
                      &= \xi^-\partial_-\tilde\varphi_+-\int_{-L-/2}^{L_-/2} dy^-\frac{\xi^-}{L_-}\partial_-\tilde\varphi_+
                        +\frac{\pi^+_++\Delta_+}{2L_-}(\xi^--\int_{-L-/2}^{L_-/2} dy^-\frac{\xi^-}{L_-}).
  \end{split}
\end{equation}
These improved transformations are canonically generated in the Dirac bracket by
the reduced Noether charge,
\begin{equation}
  \label{eq:230}
  \hat \delta^R_{\xi^-}\cdot=\{\cdot,Q^{+*}_{\xi^-}\}^*_+. 
\end{equation}
The latter form a canonical realization of the conformal algebra in the Dirac
bracket on the first class constraint surface,
\begin{equation}
  \label{eq:85}
  \{Q^{+*}_{\xi^-_1},Q^{+*}_{\xi^-_2}\}^*_+=\hat\delta^R_{\xi^-_2} Q^{+*}_{\xi^-_1}=Q^{+*}_{[\xi^-_1,\xi^-_2]}
  +\frac{(\bar g^+_+)^2}{4L_-}\int_{-L-/2}^{L_-/2} dx^-\frac{[\xi_1^-,\xi^-_2]}{L_-}. 
\end{equation}

\subsection{Preliminary attempt at quantization}
\label{sec:quantization-circle}

Even though encoding half of the relevant data in a Lagrange multiplier is fine
insofar as the characteristic initial value problem is concerned, the lack of a
symplectic structure for these data means that one does not know how to quantize
this sector. We illustrate this by providing the partition function in the
single front formulation, while treating the sector associated with the first
class constraints as pure gauge.

We consider the theory with periodic boundary conditions on the $x^+=0$ front,
$x^-\sim x^-+L_-$. Our aim here is to attempt to compute the standard partition
function,
\begin{equation}
  \label{eq:56d}
  Z(\beta,\alpha)={\rm Tr}\, e^{-\beta  \hat H +i\alpha \hat P},
\end{equation}
where the observables are the Hamiltonian and linear momentum in instant form,
from the viewpoint of canonical quantization on a single light front $x^+=c^+$.
Details on how to obtain this partition function in terms of chiral bosons in
instant form are reviewed in Appendix \ref{sec:instant-form-}.

When using that $\partial_0=\frac{1}{\sqrt 2}(\partial_++\partial_-)$,
$\partial_1=\frac{1}{\sqrt 2}(\partial_+-\partial_-)$, the generator
$Q^{+}_{\partial_0}$ of evolution in $x^0$ time and the generator
$Q^{+}_{\partial_1}$ of spatial translation in $x^1$ are
\begin{equation}
  \label{eq:54}
  \begin{split}
    & Q_{\partial_0}^+=\frac{1}{\sqrt 2}\big[\bar \lambda^+_+\bar g^+_++\bar\pi^+_+\frac{\Delta_+}{L_-}
      + \int dx^-\, (\partial_-\tilde \varphi_+)^2\big],\\
    & Q_{\partial_1}^+=\frac{1}{\sqrt 2}\big[\bar \lambda_+^{+}\bar g^+_++\bar\pi^+_+\frac{\Delta_+}{L_-}
      -\int dx^-\, (\partial_-\tilde \varphi_+)^2\big].
  \end{split}
\end{equation}

If $\tau=\frac{\alpha+i\beta}{L}$, for some $L$, the observable involved in the partition
function is
\begin{equation}
  \label{eq:55}
  -\beta Q_{\partial_0}^+-i\alpha Q_{\partial_1}^+=\frac{i\tau L}{\sqrt 2}H^{+R}-\frac{i\bar \tau L}{\sqrt 2} H^{+*},
\end{equation} 
where 
\begin{equation}
    H^{+R}= \int dx^-\, (\partial_-\tilde\varphi_+)^2,\quad H^{+*}=\bar\lambda_+^+\bar g^+_++\bar\pi^+_+\frac{\Delta_+}{L_-},
\end{equation}
so that the partition function factorizes into contributions from the chiral
sector and the zero mode,
\begin{equation}
  \label{eq:57}
  Z(\tau,\bar\tau)=Z^{+R}(\tau)Z^{+*}(\bar\tau),\quad Z^{+R}(\tau)={\rm Tr}\, e^{i\tau\frac{L}{\sqrt 2}\hat H^{+R}},\quad Z^{+*}(\bar\tau)=
  {\rm Tr}\,e^{-i\bar \tau \frac{L}{\sqrt 2} \hat H^{+*}}.
\end{equation}

The chiral field $\tilde\varphi_+(x^-)$ satisfies periodic boundary conditions in
$x^-$ of periodicity $L_-$ and can be expanded as
\begin{equation}
    \label{eq:48}
    \tilde\varphi_+(x^-)=\sum_{n>0}\frac{1}{\sqrt{2k_-L_-}}\big(a_{k_-}e^{-ik_-x^-}+{\rm c.c.}),\quad
    k_-=\frac{2\pi n}{L_-}.
  \end{equation}
In terms of the oscillators, the non-vanishing Dirac brackets
\eqref{Diracbrackets} are
\begin{equation}
  \label{eq:58}
  \{a_{k_-},a^*_{k'_-}\}^*_+=-i\delta_{n,n'},
\end{equation}
while the Hamiltonian $H^{+R}$ is
\begin{equation}
    \label{eq:65}
    H^{+R}=\frac 12 \sum_{n>0}k_-(a^*_{k_-}a_{k_-}+a_{k_-}a^*_{k_-}).
\end{equation}
When using symmetric ordering with the associated Casimir energy computed
through zeta function regularization, the quantum Hamiltonian is
\begin{equation}
    \label{eq:66}
    \hat H^{+R}=E^{+R}_0+\sum_{n>0}k_-\hat a^\dagger_{k_-}\hat a_{k_-},\quad
    E^{+R}_0=\frac{\pi}{L_-}\sum_{n>0}n=-\frac{2\pi}{24 L_-}.
\end{equation}
The associated contribution to the partition function is
\begin{equation}
    \label{eq:59}
    Z^{+R}(\tau)=\frac{1}{\eta(q(\frac{L\tau}{\sqrt 2 L_-}))},
\end{equation}
where the Dedekind eta function is defined in \eqref{eq:101}.

As is clear from Appendix \ref{sec:instant-form-}, if one wants to recover the
results for the partition function in instant form, one cannot quantize the zero
mode as a single pure gauge degree of freedom. Indeed, the associated physical
Hilbert space would then contain only the vacuum, which would lead to
$Z^{+*}(\bar\tau)=1$.

How to proceed in a way that puts right and left movers on
an equal footing from the very beginning is discussed in the next section.
Before doing that, it turns out to be useful to briefly indicate the change in
the formulas of the preceding subsection when exchanging the roles of the two
fronts.

\subsection{Results on the other front}
\label{sec:results-other-front}

At the outset, we could have chosen $x^-$ as the time and set up the problem
from the viewpoint of the $x^-=0$ front, on which we assume periodic boundary
conditions, $x^+\sim x^++L_+$. Let us just briefly indicate the associated changes.

In all formulas of the previous subsections, one merely needs to exchange all
$+$ and $-$ sub- and superscripts. In particular, the first order action
principle now becomes,
\begin{equation}
  \label{eq:7a}
  S_H[\phi,\pi^-,\lambda^-]=\int dx^-\int_{-L_+/2}^{L_+/2}dx^+\ \mathcal{L}_H^-,\quad \mathcal{L}_H^-=\pi^- \partial_-\phi
  -\lambda^-(\pi^--\partial_+\phi),
\end{equation}
while the Dirac brackets are given by
\begin{equation}
  \label{Diracbrackets1}
  \begin{split}
    & \{\tilde \varphi_-(x^+),\tilde \varphi_-(y^+)\}^*_-=-\frac{1}{4}\varepsilon(x^+-y^+)
      +\frac{x^+-y^+}{2L_+},\\
    & \{\tilde \varphi_-(x^+),\tilde \pi^-_-(y^+)\}^*_-=\frac 12 [\delta(x^+,y^+)-\frac{1}{L_+}],\\
    &  \{\tilde \pi^-_-(x^+),\tilde \pi^-_-(y^+)\}^*_-=\frac 12 \delta'(x^+,y^+),\\
    &\{\bar \varphi_-,\bar \pi^-_-\}^*_-=1.
  \end{split}
\end{equation}

The considerations of subsection \ref{sec:quantization-circle} are modified as
follows. Instead of \eqref{eq:54} and \eqref{eq:55}, one now gets
\begin{equation}
  \label{eq:110}
  \begin{split}
    & Q_{\partial_0}^-=\frac{1}{\sqrt 2}[\int dx^+(\partial_+\tilde\varphi_-)^2+\bar\lambda^-_-\bar g^-_-+\pi^-_-\frac{\Delta_-}{L_+}],\\
    & Q_{\partial_1}^-=\frac{1}{\sqrt 2}[\int dx^+(\partial_+\tilde\varphi_-)^2-\bar\lambda^-_-\bar g^-_--\pi^-_-\frac{\Delta_-}{L_+}],
  \end{split}
\end{equation}
and 
\begin{equation}
    \label{eq:55a}
    \begin{split}
      & -\beta Q_{\partial_0}^--i\alpha Q_{\partial_1}^-=i\tau\frac{L}{\sqrt 2}H^{-*}-i\bar \tau \frac{L}{\sqrt 2} H^{-L},\\
      & H^{-L}=
        \int dx^+\, (\partial_+\tilde\varphi_-)^2,\quad H^{-*}=\bar\lambda_-^-\bar g^-_-+\pi^-_-\frac{\Delta_-}{L_+},
    \end{split}
\end{equation}
while, instead of \eqref{eq:57} the partition function now becomes
\begin{equation}
  \label{eq:57a}
  Z(\tau,\bar\tau)=Z^{-*}(\tau)Z^{-L}(\bar\tau),\quad
  Z^{-*}(\tau)={\rm Tr}\, e^{i\tau\frac{L}{\sqrt 2}\hat H^{-*}},\quad Z^{-L}(\bar\tau)=
  {\rm Tr}\,e^{-i\bar \tau \frac{L}{\sqrt 2} \hat H^{-L}}.
\end{equation}

In terms of the oscillator variables in the expansion
\begin{equation}
  \label{eq:60}
  \tilde{\varphi}_-(x^+)=\sum_{n>0}\frac{1}{\sqrt{2k_+L_+}}
  \big(\tilde a_{k_+}e^{-ik_+x^+}+{\rm c.c.}),\quad k_+=\frac{2\pi n}{L_+},
 \end{equation}
the non-vanishing Dirac brackets become
\begin{equation}
  \label{eq:61}
  \{\tilde a_{k_+},\tilde a^*_{k'_+}\}^*=-i\delta_{n,n'}.
\end{equation}
In this case, one finds that $H^{-L}$ is
\begin{equation}
  \label{eq:65a}
  H^{-L}=\frac 12 \sum_{n>0}k_+(\tilde a^*_{k_+}\tilde a_{k_+}+\tilde a_{k_+}\tilde a^*_{k_+}).
\end{equation}
When using symmetric ordering with the associated Casimir energy computed
through zeta function regularization, the quantum Hamiltonian is
\begin{equation}
  \label{eq:66a}
  \hat H^{-L}=E_0^{-L}+\sum_{n>0}k_+\hat{\tilde a}^\dagger_{k_+}\hat{\tilde a}_{k_+},\quad
  E_0^{-L}=\frac{\pi}{L_+}\sum_{n>0}n=-\frac{2\pi}{24 L_+}.
\end{equation}
The associated contribution to the partition function is
\begin{equation}
  \label{eq:59a}
  Z^{-L}(\bar \tau)=\frac{1}{\widebar{\eta(q(\frac{L\tau}{\sqrt 2 L_+}))}}.
\end{equation}

The main lessons from the analysis of this section are the following

\begin{itemize}
\item {\em When setting up the canonical formalism on a single null hyperplane,
    none of the solutions or of the degrees of freedom of the system are lost.
    In the case of the $x^+$-front, the right movers and their Poisson structure
    are correctly reproduced and lead to a correct quantization of this sector.
    It turns out however that the left movers together with the momentum of the
    particle zero mode of the instant form are encoded in the Lagrange
    multiplier for a first class constraint. What is missing is a suitable
    Poisson structure that allows one to correctly quantize these degrees
    of freedom. In the case of the $x^-$ front, the role of the left and right
    movers are exchanged. }

\item {\em All the global symmetries of the system are reproduced and can be
    represented in terms of suitable conserved Noether currents. It turns out
    however that the left chiral shift symmetry is generated by a charge that
    vanishes on the constraint surface. The same applies for the charges
    generating the left chiral half of the conformal symmetries. There is no
    canonical realization for the left chiral half of the conformal symmetries.
    This is because some of the information on the full charge and the correct
    Poisson structure is missing.}
\end{itemize}

\section{Double front Hamiltonian analysis}
\label{sec:double-light-front}

We now want to recover the full phase space from a Hamiltonian viewpoint in a
symmetrical way for left and right movers. In order to do so, one sets up the
problem on two intersecting null hyperplanes, which means that one uses two
evolution directions. The union of the two phase spaces associated to these null
planes is equivalent to the phase space in instant form associated to a
space-like surface intersecting the two null ones so as to form the boundary of
a region of spacetime (see e.g.~\cite{Nagarajan1985,Mccartor1988,Hayward1993}).
The two Hamiltonians on the null planes are related to the Hamiltonian in
instant form by using Stokes' theorem. The symplectic structure on the null
planes may be described through a single covariant Poisson bracket, the Peierls
bracket
\cite{doi:10.1098/rspa.1952.0158,DeWitt:1964aa,DeWitt:1984aa,DeWitt:2003pm},
(see also
e.g.~\cite{Barnich:1991tc,Marolf:1993af,Esposito:2002ru,Khavkine:2014kya,Gieres:2021ekc}
for further considerations and reviews) that reduces to the appropriate
expressions in all cases, i.e., the usual Dirac bracket in instant form, and the
correct brackets on the two fronts.

\subsection{Covariant first order formulation adapted to front form}
\label{sec:solution-space}

A symmetric way to write the first order action
principles \eqref{eq:7} or \eqref{eq:7a} is to rename
$\lambda^{\pm}=\pi^{\mp}$, so that
\begin{equation}
    \label{eq:19}
    S_H[\phi,\pi^+,\pi^-]=\int dx^+dx^-\,\mathcal L'_H,\quad \mathcal L'_H= \pi^+\partial_+\phi+\pi^-\partial_-\phi-\pi^-\pi^+,
\end{equation}
with associated equations of motion
\begin{equation}
  \label{eq:117}
  \pi^+=\partial_-\phi,\quad \pi^-=\partial_+\phi,\quad \partial_+\pi^++\partial_-\pi^-=0.
\end{equation}
Alternatively, this can be obtained from the instant form with auxiliary fields
\eqref{eq:125} by going to light-cone coordinates $(x^+,x^-)$ when using that
$\pi^\mu$ transforms like an even vector density of weight one,
\begin{equation}
  \label{eq:109}
  \pi'^{\mu}(x')=\abs{{\rm det}(\frac{\partial x}{\partial x'})}\frac{\partial x'^\mu}{\partial x^\nu}\pi^\nu(x),\quad
  \pi^\pm= \frac{1}{\sqrt 2}(\pi^0\pm \pi^1).
\end{equation}

{\em This rewriting makes it clear that the physical interpretation of the
  Lagrange multipliers is that they are conjugate momenta for propagation along,
  rather than off the front.}

As discussed previously, if one chooses $c^-=L_-/2, c^+=L_+/2$ (as in Figure
\ref{fig:period} above), the free initial data that determines the general
solution to the equations of motion is
$\pi^+(L_+/2,x^-)$, $\pi^-(x^+,L_-/2)$, $\phi(L_+/2,L_-/2)$, and it follows from \eqref{eq:74} that
the on-shell field is given by
\begin{equation}
    \label{eq:71}
    \phi(x^+,x^-)=\phi(L_+/2,L_-/2)
    +\int_{L_-/2}^{x^-}dy^- \pi^+(L_+/2,y^-)+\int^{x^+}_{L_+/2}dy^+ \pi^-(y^+,L_-/2).
\end{equation}

\subsection{Conserved currents and Stokes' theorem}
\label{sec:curr-stok-theor}

In order to relate to the well-understood principles in instant form, one uses
Stokes' theorem on the level of conserved currents. Indeed, consider a conserved
current $j^\mu$ with conservation equation $\partial _\mu j^\mu\approx 0$, where the weak equality
now denotes an equality that holds when the equations of motion (and their
linearization) are satisfied. Defining the associated $n-1$-form on Minkowsi
spacetime by
\begin{equation}
  \label{eq:67}
  j=d^{n-1}x_\mu j^\mu.
\end{equation}
with 
\begin{equation}
  \label{eq:111}
  d^{n-k}x_{\mu_1\dots\mu_k}=\frac{1}{(n-k)!}\epsilon_{\mu_1\dots \mu_k\nu_{k+1}\dots\nu_n}dx^{\nu_{k+1}}\dots dx^{\nu_n},
\end{equation}
with $\epsilon_{01\dots n-1}=1$, $\epsilon_{\mu_0\dots \mu_{n-1}}$ completely
skew-symmetric and the wegde product being understood, the current is a vector
density of weight one,
\begin{equation}
  \label{eq:69}
  j'^{\mu}(x')={\rm det}(\frac{\partial x}{\partial x'})\frac{\partial x'^\mu}{\partial x^\nu}j^\nu(x).
\end{equation}
It then follows from Stokes' theorem that
\begin{equation}
  \label{eq:64}
  \int_{\partial V} j=\int_V dj\approx 0.
\end{equation}
In our case
\begin{equation}
  \label{eq:108}
  \begin{split}
    j&=dx^1j^0-dx^0j^1=dx^-j^+-dx^+j^-,\quad j^\pm=-\frac{1}{\sqrt 2}(j^0\pm j^1),\\
    dj&=dx^0dx^1 (\partial_0j^0+\partial_1j^1)=dx^+dx^-(\partial_+j^++\partial_-j^-).
  \end{split}
\end{equation}
In instant front, the relevant conserved quantity is
$\int_{-L/2}^{L/2} dx^1 j^0|_{x^0=0}$. Using Stokes' theorem for the closed
contour given by the red and blue lines in figure \ref{fig:period} above, it is
given by
\begin{equation}
    \label{eq:53}
       \int_{-L/2}^{L/2} dx^1 j^0|_{x^0=0}\approx-\int_{-L_+/2}^{L_+/2}dx^+j^-|_{x^-=L_-/2}-\int_{-L_-/2}^{L_-/2}dx^- j^+|_{x^+=L_+/2},
\end{equation}
In the particular case when $\partial_+ j^+\approx 0\approx \partial_-j^-$, the integrals on the right hand side
may be evaluated at any constant line $x^\pm=c^\pm$.

\subsection{Conserved symplectic $(2,n-1)$ form and Peierls bracket}
\label{sec:symplectic-2-n}

In order to be able to deal with Poisson brackets, one may consider (odd) field
variations $d_V=d_V\phi^i\frac{\partial}{\partial \phi^i}+\partial_\mu d_V\phi^i\frac{\partial}{\partial (\partial_\mu\phi^i))}+\dots$
that anti-commute with the total differential $d_H=dx^\mu\partial_\mu$,
$\partial_\mu=\frac{\partial}{\partial x^\mu}+\partial_\mu\phi^i\frac{\partial}{\partial \phi^i}+\dots $ in the context of the
variational bicomplex (see e.g.~\cite{Andersonbook,Anderson1991,Olver:1993}).

The symplectic potential $a$ is the $(1,n-1)$ form necessary to make the
Euler-Lagrange derivatives appear in the first variational formula through
``integrations by parts'',
\begin{equation}
  \label{eq:113}
  d^nx d_V\mathcal L=d^nx d_V\phi^i\frac{\delta \mathcal L}{\delta\phi^i}+d_H a,\quad a=d^{n-1}x_\mu a^\mu.
\end{equation}
The second variational formula is obtained by applying $d_V$ again,
\begin{equation}
  \label{eq:115}
  0=-d^nxd_V\phi^id_V\frac{\delta \mathcal L}{\delta\phi^i}+d_H \sigma,\quad \sigma=(-)^{n-1}d_Va=d^{n-1}x_\mu d_Va^\mu,
\end{equation}
so that the symplectic $(2,n-1)$ form $\sigma$ is conserved when the linearized
equations of motion hold,
\begin{equation}
  \label{eq:116}
  d_H\sigma\approx 0.
\end{equation}

In our case of action $\eqref{eq:19}$, the symplectic potential $a$ and
$(2,1)$ form $\sigma$ are given by
\begin{equation}
  \label{eq:70}
  a=dx^-\pi^+d_V\phi -dx^+ \pi^-d_V\phi,\quad \sigma=dx^- d_V\pi^+ d_V\phi -dx^+ d_V\pi^-d_V\phi.
\end{equation}
Note that $\sigma^+=d_V\pi^+d_V\phi,\sigma^-=d_V\pi^-d_V\phi$ do not satisfy
$\partial_+\sigma^+\approx 0\approx \partial_-\sigma^-$ so that one may not change
arbitrarily the location of the initial value null lines when applying Stokes'
theorem as in \eqref{eq:53}. One may however improve the symplectic current
through the divergence of a superpotential,
\begin{equation}
  \label{eq:200}
  \begin{split}
  & \sigma'^+=\sigma^+-\partial_-k^{+-},\quad \sigma'^-=\sigma^-+\partial_+ k^{+-},\\
  & k^{+-}=\frac 14 \int dy^+\int dy^- \varepsilon(x^+-y^+)\varepsilon(x^--y^-)d_V\pi^+(y^+,y^-)d_V\pi^-(y^+,y^-),
\end{split}
\end{equation}
so that $\partial_+\sigma'^+\approx 0\approx \partial_-\sigma'^-$ and the locations may be changed arbitrarily. Note
that this singular superpotential, when written in terms of the Heaviside step
function,
$\varepsilon(x^+-y^+)\varepsilon(x^--y^-)=4\theta(x^+-y^+)\theta(x^--y^-)-2$, is
somewhat reminiscent of the magnetic field associated with a Dirac string.

For a conserved current $\partial_+j^++\partial_-j^-\approx 0$, it follows from the discussion of
Stokes' theorem in \eqref{eq:53} that, when working at fixed $x^+$, the
generator $\int_{-L/2}^{L/2}dx^1 j^0|_{x^0=0}$ that can be directly compared to
the one in instant form, is the sum of
$-\int^{L_-/2}_{-L_-/2}dx^- j^+|_{x^+=L_+/2}$ and of
$-\int^{L_+/2}_{-L_+/2}dx^+ j^-|_{x^-=L_-/2}$ when $j^\mu$ transforms like a
vector density of weight one. Note however that, because $\pi^\mu$ transforms
like an even vector density of weight one, in the case of $\sigma'^\mu$, this
becomes the sum of $\int^{L_-/2}_{-L_-/2}dx^- \sigma'^+$ and of
$\int^{L_+/2}_{-L_+/2}dx^+ \sigma'^-$ evaluated at any constant $x^+$,
respectively $x^-$ front. The inverse of the covariant symplectic two-form is
the Peierls bracket, which preserves the ideal of functions that vanish on shell
since it gives zero when applied to equations of motion. 

When working on two-dimensional Minkowski spacetime with Fourier transforms,
where particle zero modes do not play a role, it follows from \eqref{eq:190b}
that the fundamental Peierls brackets are given by
\begin{equation}
  \label{eq:196}
  \{\phi(x^+,x^-),\phi(y^+,y^-)\}=-\frac 14 \varepsilon(x^+-y^+)- \frac 14 \varepsilon(x^--y^-).
\end{equation}
It follows in particular that 
\begin{equation}
  \label{eq:159}
  \begin{split}
    &\{\phi(x^+,x^-),\phi(x^+,y^-)\}=-\frac 14 \varepsilon(x^--y^-),\\
    & \{\phi(x^+,x^-),\phi(y^+,x^-)\}=-\frac 14 \varepsilon(x^+-y^+),\\
    & \{\phi(x^+,x^-),\pi^+(x^+,y^-)\}=\frac 12 \delta(x^-,y^-),\\
    & \{\phi(x^+,x^-),\pi^-(y^+,x^-)\}=\frac 12 \delta(x^+,y^+), \\
    &  \{\phi(x^+,x^-),\pi^-(x^+,y^-)\}=0=\{\phi(x^+,x^-),\pi^+(y^+,x^-)\},\\
    &\{\pi^+(x^+,x^-),\pi^+(x^+,y^-)\}=\frac 12 \delta'(x^-,y^-),\\
    &\{\pi^-(x^+,x^-),\pi^-(y^+,x^-)\}=\frac 12
      \delta'(x^+,y^+),\\
    & \{\pi^-(x^+,x^-),\pi^+(x^+,y^-)\}=0=\{\pi^+(x^+,x^-),\pi^-(y^+,x^-)\}_.
    \end{split}
\end{equation}

\subsection{Shift and conformal symmetries}
\label{sec:dyn1}

Stokes' theorem as in \eqref{eq:53}, suitably adjusted for the transformation of
$j^\mu_{\epsilon^\pm}$ gives for the Noether charges associated to the shift
symmetries
\begin{equation}
  \label{eq:173}
  Q_{\epsilon^\pm}\approx Q^-_{\epsilon^\pm}+Q^+_{\epsilon^\pm},\ Q^-_{\epsilon^\pm}=\int dx^+ j^-_{\epsilon^\pm}|_{x^-=L_-/2},\ Q^+_{\epsilon^\pm}=\int dx^- j^+_{\epsilon^\pm}|_{x^+=L_+/2},
\end{equation}
with
\begin{equation}
  \label{eq:174}
     j^+_{\epsilon^+}=\pi^+\epsilon^+,\ j^-_{\epsilon^+}=\pi^-\epsilon^+-\phi\partial_+\epsilon^+,\quad j^+_{\epsilon^-}=\pi^+\epsilon^--\phi\partial_-\epsilon^-,\ j^-_{\epsilon^-}=\pi^-\epsilon^-.
 \end{equation}
These expressions for the currents can either be obtained directly from
\eqref{eq:177a} or from \eqref{eq:21}, \eqref{eq:N2s}, after renaming
$\lambda^+=\pi^-$. With the brackets as in \eqref{eq:159}, it follows however
that the charges $Q_{\epsilon^\pm}$ do not generate the correct shift
symmetries in the Peierls bracket.
The ones that do are the ones for which the currents differ by
a superpotential, as in \eqref{eq:super}, \eqref{eq:46}, and for which the charges
on the two fronts are given by
\begin{equation}
  \label{eq:188}
  \begin{split}
    & Q^+_{\epsilon^+}=\int dx^- (\pi^+-\partial_-\phi)\epsilon^+\approx 0,\\
    & Q^-_{\epsilon^+}=\int dx^+ (\pi^-+\partial_+\phi)\epsilon^+\approx \int dx^+ 2\partial_+\phi \epsilon^+,\\
    & Q^+_{\epsilon^-}=\int dx^- (\pi^++\partial_-\phi)\epsilon^-\approx \int dx^- 2\partial_-\phi \epsilon^-,\\
    &  Q^-_{\epsilon^-}=\int dx^+ (\pi^--\partial_+\phi)\epsilon^-\approx 0.
  \end{split}
\end{equation}
Since the Peierls bracket of equations of motions vanishes, the shift symmetry
\begin{equation}
  \label{eq:198}
  \delta_{\epsilon^+}\phi=\epsilon^+,\quad \delta_{\epsilon^+}\pi^+=0,\quad \delta_{\epsilon^+}\pi^-=\partial_+\epsilon^+,
\end{equation}
is entirely generated by $Q^-_{\epsilon^+}$ on the constant $x^-$ front while
\begin{equation}
  \label{eq:199}
  \delta_{\epsilon^-}\phi=\epsilon^-,\quad \delta_{\epsilon^-}\pi^+=\partial_-\epsilon^-,\quad \delta_{\epsilon^-}\pi^-=0,
\end{equation}
is entirely generated by $Q^+_{\epsilon^-}$ on the constant $x^+$ front.

Similarly, the charges $Q'_{\xi}=Q'^-_{\xi}+Q'^+_{\xi}$ associated to the improved
currents
\begin{equation}
  \label{eq:201}
   j'^+_\xi=\xi^-\pi^+\partial_-\phi-\xi^+\pi^-(\partial_-\phi-\pi^+),\quad j'^-_\xi=\xi^+\pi^-\partial_+\phi-\xi^-\pi^+(\partial_+\phi-\pi^-).
\end{equation}
(see \eqref{eq:168} or \eqref{eq:33}, \eqref{eq:37}) for conformal symmetries,
depend on-shell only on the right and left chiral halves, 
\begin{equation}
  \label{eq:202}
  Q'^+_{\xi}\approx \int dx^- \xi^-(\partial_-\phi)^2=Q'^+_{\xi^-},\quad Q'^-_{\xi}\approx \int dx^+ \xi^+(\partial_+\phi)^2=Q'_{\xi^+}.
\end{equation}
They correctly generate these halves of the conformal symmetries in the Peierls
bracket,
\begin{equation}
  \label{eq:187}
  \{\phi,Q'^+_{\xi^-}\}=\xi^-\partial_-\phi,\quad \{\phi,Q'^-_{\xi^+}\}=\xi^+\partial_+\phi,
\end{equation}
and represent the conformal algebra on-shell,
\begin{equation}
  \label{eq:203}
  \{Q'^+_{\xi_1^-},Q'^+_{\xi_2^-}\}= Q'^+_{[\xi_1^-,\xi_2^-]},\quad  \{Q'^-_{\xi_1^+},Q'^-_{\xi_2^+}\}= Q'^-_{[\xi_1^+,\xi_2^+]},\quad \{Q'^+_{\xi_1^-},Q'^-_{\xi_2^+}\}\approx 0,
\end{equation}
provided one may integrate by parts both in $x^-$ and in $x^+$.

After the usual quantization of the left and right movers on Minkowski space,
the computation of the partition function at large volume/ high temperature,
which involves adding a factor $L/2\pi$ to the momentum space integral, gives the
scalar black body result, $\ln Z(\beta)=\frac{\pi}{6}\frac{L}{\beta}$, which is the large
volume/ high temperature limit of the result with finite size corrections that we
are going to discuss in the next section. 

\subsection{Time-like cylinder}
\label{sec:part-funct-large}

\subsubsection{Chiral boson and zero mode sectors}
\label{sec:chiral-boson-zero}

More interesting is to consider the system on the standard time-like cylinder.
Adapted to this topology, one may split the fields into their zero modes and the
rest as in \eqref{eq:63},
\begin{equation}
  \label{eq:210}
  \phi=\bar\phi_0+ \tilde\phi_0,\quad \pi^\mu=\frac 1L\bar\pi^\mu_0+\tilde\pi^\mu_0.
\end{equation}
Note that one may also extract zero modes on the two fronts. The lower $\pm$ index
indicates extracting at fixed $x^\pm$ the $x^\mp$ zero mode. Upper and lower indices
are uncorrelated:
\begin{equation}
  \label{eq:118}
  \pi^\pm(x^+,x^-)=\frac{1}{L_\mp}\bar\pi^\pm_\pm(x^\pm)+\tilde\pi^\pm_\pm(x^+,x^-),
  \quad\phi(x^+,x^-)=\bar\phi_\pm(x^\pm)+\tilde\phi_\pm(x^+,x^-),
\end{equation}
with $\bar\pi^\pm_\pm(x^\pm)=\int_{-L_\mp/2}^{L_\mp/2} dx^\mp \pi^\pm_\pm(x^+,x^-)$,
$\bar\phi_\pm(x^\pm)=\frac{1}{L_{\mp}}\int_{-L_\mp/2}^{L_\mp/2} dx^\mp \phi(x^+,x^-)$.
In terms of the new notation, $\bar\lambda^+_+(x^+)=\frac{1}{L_-}\bar\pi^-_+(x^+)$,
$\bar\lambda^-_-(x^-)=\frac{1}{L_+}\bar\pi^+_-(x^-)$. 
Note however that the zero modes are not independent, since by definition,
\begin{equation}
  \label{eq:137}
  \begin{split}
     \int^{L_{+}/2}_{-L_+/2}dx^{+}\bar\pi^{+}_{+}(x^{+})&= \int^{L_{-}/2}_{-L_-/2}dx^{-}\bar\pi^{+}_{-}(x^{-}),\\
      \int^{L_{-}/2}_{-L_-/2}dx^{-}\bar\pi^{-}_{-}(x^{-})&= \int^{L_{+}/2}_{-L_+/2}dx^{+}\bar\pi^{-}_{+}(x^{+}),\\
     \frac{1}{L_{+}}\int^{L_{+}/2}_{-L_+/2}dx^{+}\bar\phi_{+}(x^{+})&=
      \frac{1}{L_{-}}\int^{L_{-}/2}_{-L_{-/2}}dx^{-}\bar\phi_{-}(x^{-}).
  \end{split}
\end{equation}

We begin by writing the on-shell field \eqref{eq:71} as
\begin{multline}
    \label{eq:129}
    \phi(x^+,x^-)=\phi(0,0)+\frac{x^-\bar\pi^+_+(L_+/2)}{L_-}+\frac{x^+\bar\pi^-_-(L_-/2)}{L_+}\\
    +\int_0^{x^-}dy^-  \tilde\pi^+_+(L_+/2,y^-)+\int^{x^+}_{0}dy^+ \tilde\pi^-_-(y^+,L_-/2).
  \end{multline}
In this on-shell context, we may define the chiral fields as the primitives
\begin{equation}
  \phi^R(x^-)=\partial_-^{-1}\tilde \pi^+_+(L_+/2,x^-),\quad \phi^L(x^+)=\partial_+^{-1}\tilde \pi^-_-(x^+,L_-/2)
\end{equation}
that do not contain zero modes, or explicitly,
\begin{equation}
  \begin{split}
    \phi^R(x^-)&=\int_0^{x^-}dy^-  \tilde\pi^+_+(L_+/2,y^-)-\frac{1}{L_-}\int_{-L_-/2}^{L_-/2}dy^-\int_0^{y^-}dz^-  \tilde\pi^+_+(L_+/2,z^-),\\
    \phi^L(x^+)&=\int_0^{x^+}dy^+  \tilde\pi^-_-(y^+,L_-/2)-\frac{1}{L_+}\int_{-L_+/2}^{L_+/2}dy^+\int_0^{y^+}dz^+  \tilde\pi^+_+(z^+,L_-/2). 
  \end{split}
\end{equation}
One may also apply Stokes' theorem in the form of \eqref{eq:53} to the $\pi^\mu$
themselves which form a conserved current according to \eqref{eq:117},
\eqref{eq:128}, and satisfy in addition
$\partial_+\pi^+\approx0\approx \partial_-\pi^-$. When taking into account that
they transform like an even vector density of weight one, it follows that
\begin{equation}
    \label{eq:75}
    \bar\pi^0_0(0)\approx \bar\pi^-_-({L_-}/{2})+\bar\pi^+_+({L_+}/{2})\approx \bar\pi^-_-(0)+\bar\pi^+_+(0),
\end{equation}
and that the on-shell field can be written as
\begin{equation}
    \label{eq:131}
    \phi(x^{+},x^{-})=\phi(0,0)-\phi^{R}(0)-\phi^{L}(0)+\frac{x^-\bar\pi^+_+(0)}{L_-}+\frac{x^+\bar\pi^-_-(0)}{L_+}
    +\phi^{R}(x^{-})+\phi^{L}(x^{+}).
\end{equation}
The periodicity condition \eqref{eq:119} then requires the matching condition
\begin{equation}
    \label{eq:134}
    \bar\pi^+_+(0)=\bar\pi^-_-(0).
\end{equation}
On account of \eqref{eq:75} this implies that
\begin{equation}
    \label{eq:add}
    \bar\pi^+_+(0)=\bar\pi^-_-(0)=\frac 12\bar\pi^{0}_{0}(0).
  \end{equation}

By integrating over $x^{1}$ at $x^{0}=0$, one then finds that
\begin{equation}
  \label{eq:132}
  \bar\phi_{0}(0)=\phi(0,0)-\phi^{R}(0)-\phi^{L}(0),
\end{equation}
so that \eqref{eq:131} becomes
\begin{equation}
  \label{eq:133}
  \phi(x^{+},x^{-})=\bar\phi_{0}(0)+(\frac{x^{+}+x^-}{\sqrt 2L})\bar\pi^0_0(0)
  +\phi^{R}(x^{-})+\phi^{L}(x^{+}),
\end{equation}
in agreement with the solution in instant form \eqref{eq:114}.

\subsubsection{Peierls brackets}
\label{sec:peierls-brackets}

By performing first the integration of $x^1$, the action \eqref{eq:19}, or
equivalently \eqref{eq:125}, splits into 
\begin{equation}
  \label{eq:205}
  S'_H[\phi,\pi^\mu]=S'_P[\bar\phi_0,\bar\pi^\mu_0]+S'_H[\tilde\phi_0,\tilde\pi^\mu_0], 
\end{equation}
with a particle action given by 
\begin{equation}
  \label{eq:209}
  S'_P[\bar\phi_0,\bar\pi^\mu_0]=\int dx^0[\bar\pi^0_0\partial_0\bar\phi_0-\frac{1}{2L}(\bar\pi^0_0)^2+\frac{1}{2L}(\bar\pi^{1}_0)^2].
\end{equation}
The associated general solution is $\bar\pi^1_0=0$, $\bar\pi^0_0=\bar\pi^0_0(0)$
$\bar\phi_0=\bar\phi_0(0)+\frac{x^0}{L}\bar\pi^0_0(0)$, and the Peierls bracket
is given by
\begin{equation}
  \label{eq:207}
  \{\bar\phi_0(x^0),\bar\phi_0(y^0)\}=-\frac{x^0-y^0}{L}.
\end{equation}

How to deal with $S'_H[\tilde\phi_0,\tilde\pi^\mu_0]$ in terms of chiral fields in instant
form is reviewed in detail in appendix \ref{sec:decomposition-cylinder} and
\ref{sec:mode-expans-part}. Here, we may proceed as follows. It follows from
\eqref{eq:159} and \eqref{eq:207} that
\begin{multline}
  \label{eq:204}
  \{\tilde\phi_0(x^+,x^-),\tilde\phi_0(y^+,y^-)\}=-\frac 14 \varepsilon(x^0-x^1-y^0+y^1)-\frac 14
  \varepsilon(x^0+x^1-y^0-y^1)+\frac{x^0-y^0}{L}\\
  =-\frac 14 \varepsilon(x^--y^-)-\frac 14 \varepsilon(x^+-y^+)+\frac{x^--y^-}{2L_-}
  +\frac{x^+-y^+}{2L_+}.
\end{multline}
On account of \eqref{eq:133}, the solution $\tilde\phi_0$ decomposes as 
\begin{equation}
  \label{eq:208}
  \tilde\phi_0=\phi^R(x^-)+\phi^L(x^+),
\end{equation}
for periodic functions $\phi^R(x^-),\phi^L(x^+)$ that do not contain an $x^-$, respectively
an $x^+$, zero mode. By differentiating \eqref{eq:204} and using \eqref{eq:208},
it follows that
\begin{equation}
  \label{eq:212}
  \begin{split}
    \{\partial_{x^-}\phi^R(x^-),\partial_{y^-}\phi^R(y^-)\}&=\frac 12 \delta'(x^-,y^-),\\
    \{\partial_{x^+}\phi^L(x^+),\partial_{y^+}\phi^L(y^+)\}&=\frac 12 \delta'(x^+,y^+),\\
    \{\partial_{x^-}\phi^R(x^-),\partial_{y^+}\phi^L(y^+)\}&=0.
  \end{split}
\end{equation}
Following the same reasoning as in instant form in \eqref{eq:103}, this implies
that
\begin{equation}
  \label{eq:211}
  \begin{split}
    \{\phi^R(x^-),\phi^R(y^-)\}&=-\frac 14 \varepsilon(x^--y^-)+\frac{x^--y^-}{2L_-}
    =-\sum_{n>0}\frac{1}{2\pi n}\sin{\frac{2\pi n}{L_-}(x^--y^-)},\\
    \{\phi^L(x^+),\phi^L(y^+)\}&=-\frac 14 \varepsilon(x^+-y^+)+\frac{x^+-y^+}{2L_+}
    =-\sum_{n>0}\frac{1}{2\pi n}\sin{\frac{2\pi n}{L_+}(x^+-y^+)},\\
    \{\phi^R(x^-),\phi^L(x^+)\}&=0.
  \end{split}
\end{equation}

\subsubsection{Partition function}
\label{sec:part-funct}

Since the Dirac brackets of $\tilde\phi_+(x^-),\tilde\phi_-(x^+)$ in
\eqref{Diracbrackets}, \eqref{Diracbrackets1} are the same as the Peierls
brackets of $\phi^R(x^-),\phi^L(x^-)$, it follows that if one expands the latter
in terms of modes in the same way as the former in equations \eqref{eq:48},
\eqref{eq:60}, the Peierls bracket of the oscillators are identical to the Dirac
brackets in \eqref{eq:58}, \eqref{eq:61}.

When using the on-shell field in \eqref{eq:133}, the generators for shift and conformal
symmetries in \eqref{eq:188}, \eqref{eq:202} decompose as 
\begin{equation}
  \label{eq:213}
  \begin{split}
    Q^-_{\epsilon^+}&=\frac{\sqrt 2\bar\pi^0_0}{L}\int dx^+\epsilon^++2\int dx^+\partial_+\phi^L\epsilon^+,\\
    Q^+_{\epsilon^-}&=\frac{\sqrt 2\bar\pi^0_0}{L}\int dx^-\epsilon^-+2\int dx^-\partial_-\phi^R\epsilon^-,\\
  Q'_{\xi^-}&=\frac{(\bar\pi^0_0)^2}{2L^2}\int dx^-\xi^-+\frac{\sqrt 2\bar\pi^0_0}{L}\int dx^-\xi^-\partial_-\phi^R+\int dx^-\xi^-(\partial_-\phi^R)^2,\\
    Q'_{\xi^+}&=\frac{(\bar\pi^0_0)^2}{2L^2}\int dx^+\xi^++\frac{\sqrt 2\bar\pi^0_0}{L}\int dx^+\xi^+\partial_+\phi^L+\int dx^+\xi^+(\partial_+\phi^L)^2,
  \end{split}
\end{equation}
where all integrals are over one period. 
This implies that the generator for time evolution is given by
\begin{equation}
  \label{eq:214}
  Q'_{\partial_0}=Q'_{\xi^-=1/\sqrt
    2}+ Q'_{\xi^+=1/\sqrt 2}=\frac{(\bar\pi^0_0)^2}{2L}+\frac{1}{\sqrt 2}\int dx^-(\partial_-\phi^R)^2+\frac{1}{\sqrt 2}\int dx^+(\partial_+\phi^L)^2, 
\end{equation}
while the generator for space evolution is given by
\begin{equation}
  \label{eq:215}
  Q'_{\partial_1}=Q'_{\xi^-=-1/\sqrt
    2}+ Q'_{\xi^+=1/\sqrt 2}=-\frac{1}{\sqrt 2}\int dx^-(\partial_-\phi^R)^2+\frac{1}{\sqrt 2}\int dx^+(\partial_+\phi^L)^2.
\end{equation}
Since the particle sector and its contribution to $Q'_{\partial_0}$ is the same as in
instant form, as shown in appendix \ref{sec:mode-expans-part}, its contribution to the partition function is given by \eqref{eq:102}.
Furthermore, since the contributions of the chiral sectors agree with the
contributions in the chiral sectors of $Q^+_{\partial_0},Q^-_{\partial_0}$ in
\eqref{eq:54}, \eqref{eq:110}, their contribution to the partition function
$Z(\tau,\bar\tau)$ are given in \eqref{eq:59}, \eqref{eq:59a} and reduce to the
two chiral contributions $Z^R(\tau)Z^L(\bar\tau)$ in instant form given in
\eqref{eq:100} when taking into account that $L_{\pm}=\frac{L}{\sqrt 2}$. We
thus end up with the well-known modular invariant result
\begin{equation}
  \label{eq:63a}
  Z(\tau,\bar\tau)= \frac{1}{\sqrt{\tau_2}|\eta (\tau)|^2}.
\end{equation}

\subsection{Alternative double front constrained Hamiltonian analysis}
\label{sec:altern-double-front}

Even though the exact Peierls bracket for free field theories is very appealing,
it is possible to formulate all results consistently from the knowlegde
of the brackets on equal ``time'' slices, because the Peierls bracket is
uniquely determined by suitable initial conditions and the fact that it is a
solution to the homogeneous equations of motion (cf.~\eqref{eq:189}). In
addition, such a formulation has the advantage that it extends to interacting
theories where an exact, non-perturbative solution of the equations of motion is
usually not readily accessible.

One may now develop in parallel the constrained Hamiltonian analysis on the two
intersecting fronts $x^+=L_+/2$ and $x^-=L_-/2$ following what has been done for
single fronts in the previous section. We will ``match'' two equivalent
off-shell description of the theory, i.e., two different expressions for the
phase spaces and their dynamics rather than merely matching two different
expressions for solutions, as done for instance in equation \eqref{eq:74}.

More concretely, the system described by action \eqref{eq:125} has two
equivalent expressions for a constrained Hamiltonian system with $\pi^\mp$, the
Lagrange multipliers for the constraint $g^\pm=\pi^\pm-\partial_\mp\phi$. In each of those
expressions, one uses different variables with zero modes at $x^+=L_+/2$ and
$x^-=L_-/2$ extracted,
\begin{equation}
    \label{eq:122}
    \begin{split}
      S_{H}^{I+}&=\int_{-\infty}^{+\infty} dx^+\big[2\bar L^+_H
               +\int_{x^+-L_-}^{x^++L_-} dx^-[\tilde\pi^+_+\partial_+\tilde\varphi_+-\tilde\pi^-_+\tilde g^+_+]\big],\\
      S_{H}^{I-}&=\int_{-\infty}^{+\infty} dx^-\big[2\bar L^-_H
             +\int_{x^--L_+}^{x^-+L_+}dx^+[\tilde\pi^-_-\partial_-\tilde\varphi_--\tilde\pi^+_-\tilde g^-_-]\big].
    \end{split}
  \end{equation}
  Note that in the first expression, one integrates over the standard cylinder
  by cutting with constant $x^+$ slices, with $x^+\in ]-\infty,+\infty[$, and
  $x^-\in[x^+-L_-,x^++L_-]$, whereas is in the second expression, one uses
  constant $x^-$ slices, with $x^-\in ]-\infty,+\infty[$, and $x^+\in[x^--L_+,x^-+L_+]$, As a
  consequence, since each of these actions represents the initial action, it is
  also represented by their sum divided by two,
\begin{equation}
  \label{eq:130}
  S^{I}_H=\frac{1}{2} S_{H}^{I+}
  +\frac 12 S_{H}^{I-}.
\end{equation}
Regrouping terms differently, and using the periodicity of the fields, we may
also write
\begin{equation}
  \label{eq:135}
  S^{I}_H=\bar S+
  \tilde S^R+\tilde S^L,
\end{equation}
where
\begin{equation}
  \label{eq:136}
  \begin{split}
    & \tilde S^R=\int_{-\infty}^{+\infty} dx^+\int_{-L_-/2}^{L_-/2} dx^-[\tilde\pi^+_+\partial_+\tilde\varphi_+-\tilde\pi^-_+\tilde g^+_+],\\
    & \tilde S^L=\int_{-\infty}^{+\infty} dx^-\int_{-L_+/2}^{L_+/2} dx^+[\tilde\pi^-_-\partial_-\tilde\varphi_--\tilde\pi^+_-\tilde g^-_-],
  \end{split}
\end{equation}
and
\begin{multline}
  \label{eq:87}
  \bar S= \int_{-\infty}^{+\infty} dx^+[\bar\pi^+_+\partial_+\bar \varphi_+-\frac{1}{L_-}\bar\pi^-_+(\bar\pi^+_+-\Delta_+)+\bar \pi^+_+\frac{\Delta_+}{L_-}-2\frac{\Delta^2_+}{L_+}]
  \\ + \int_{-\infty}^{+\infty} dx^-[\bar\pi^-_-\partial_-\bar \varphi_--\frac{1}{L_+}\bar\pi^+_-(\bar\pi^-_--\Delta_-)+\bar
  \pi^-_-\frac{\Delta_-}{L_+}
  -2\frac{\Delta^2_-}{L_-}].
\end{multline}

The analysis of $\tilde S^R,\tilde S^L$ then proceeds as in section
  \ref{sec:light-cone-hamilt}. In particular, since $L_\pm=\frac{L}{\sqrt 2}$,
  the expansions of the chiral fields \eqref{eq:48}, \eqref{eq:60} agree with
  the on-shell expansions in instant form \eqref{eq:94},
\begin{equation}
  \label{eq:112}
  \tilde\phi_+(x^-)=\phi^R(x^{-}),\quad \tilde \phi_-(x^+)=\phi^L(x^+).
\end{equation}
The contribution of the chiral sectors described by $\tilde S^{R},\tilde S^{L}$ to
the partition function $Z(\tau,\bar\tau)$ are given in \eqref{eq:59} and
\eqref{eq:59a} and reduce to the two chiral contributions
$Z^R(\tau)Z^L(\bar\tau)$ in instant form given in \eqref{eq:100} since
$L_{\pm}=\frac{L}{\sqrt 2}$.

The sector described by action \eqref{eq:87} is more subtle.
First there are the dynamically imposed matching conditions corresponding to
\eqref{eq:223}, which can be written on-shell as
$\Delta_+=\bar{\bar\pi}^-_+$, $\Delta_-=\bar{\bar\pi}^+_-$. 
Furthermore, the variables are constrained by the integral relations
\eqref{eq:137}. One way to proceed so as to get the correct result is as
follows. In order not to over-count left and right movers, we use the procedure
discussed at the end of section \ref{sec:impr-acti-princ}, and make $\Delta_+,\Delta_-$ 
dynamical fields. After solving the constraints in the action, one ends up with
\begin{equation}
  \label{eq:139}
  S^P=\int_{-\infty}^{+\infty} dx^+[\bar\pi^+_+\partial_+\bar\varphi_+-\frac{(\bar\pi^+_+)^2}{L_+}]
  +\int_{-\infty}^{+\infty} dx^-[\bar\pi^-_-\partial_-\bar\varphi_+-\frac{(\bar\pi^-_-)^2}{L_-}]. 
\end{equation}
After the change of integration variables $x^0=\sqrt 2 x^+$ and $x^0=\sqrt 2
x^-$ in the two integrals, and imposing off-shell the matching conditions
\begin{equation}
  \label{eq:140}
  \bar\pi^+_+=\bar\pi^-_-=\frac 12 \bar\pi^0_0,\quad \bar\varphi_+=\bar\varphi_-=\bar\phi_0,
\end{equation}
in the action, one ends up with the standard zero mode particle action
\eqref{eq:126} of the instant form. The associated contribution to the partition
function then follows as in the appendix.

The lessons that can drawn from the double front analysis are:

\begin{itemize}
\item {\em The missing part of the symplectic structure is determined by equal time
    Poisson brackets on the other front. The left chiral shifts and the left
    half of the conformal algebra are entirely generated by non-vanishing
    charges on the other front.}

\item {\em In the case of the time-like cylinder, the $x^1$ zero mode may be
    extracted at the start of the computation. If one works on-shell with
    Peierls brackets, the symplectic structure of the particle zero mode of
    the instant form is automatically taken into account and there are no longer
    any $x^-$, $x^+$ zero modes left to worry about.}

\item {\em If one works separately on each of the fronts, due care has to be taken
    not to over-count when putting the sectors back together.}
 
\end{itemize}

\section{Massive case}
\label{sec:dyn2}

We now illustrate how some of the subtleties related to zero modes are absent if one adds a mass term.

\subsection{Dynamics, symmetries and Noether currents}
\label{sec:dynamics-2}

The action is now
\begin{equation}
  \label{eq:LM}
  S= \int dx^{+}dx^{-}[\partial_{+}\phi\partial_{-}\phi-\frac 12 m^{2}\phi^{2}].
\end{equation}
There are no longer any shift symmetries while the conformal symmetries reduce
to the Poincar\'e symmetries in $1+1$ spacetime dimensions. The latter are
described by $\delta_{\xi}\phi=\xi^{\rho}\partial_{\rho}\phi$ where
\begin{equation}
  \xi^{+}=a^{+}+\omega x^{+},\quad \xi^{-}=a^{-}-\omega x^{-},
\end{equation}
or equivalently, by $\partial_-\xi^+=0=\partial_+\xi^-$ and in addition
$\partial_+\xi^++\partial_-\xi^-=0$. The associated Noether current is given in
terms of the trace-full energy momentum tensor
\begin{equation}
  \label{eq:146}
  T_{\pm\pm}=(\partial_\pm\phi)^2,\quad
  T_{\mp\pm}=\frac{m^2}{2}\phi^2,
\end{equation}
as
\begin{equation}
  j^+_{L,\xi}=T_{--}\xi^-+T_{-+}\xi^+,\quad
  j^-_{L,\xi}=T_{++}\xi^++T_{+-}\xi^-.
\end{equation}

\subsection{Single front Hamiltonian analysis}
\label{sec:hamiltonian-analysis}

In the canonical analysis on the $x^{+}=c^{+}$ front, the constraint
\eqref{eq:5} is unchanged but the primary Hamiltonian is no longer proportional
to this constraint, but rather
\begin{equation}
  \label{eq:144}
  H=\int_{-L_-/2}^{{L_{-}/2}} dx^{-}\frac{m^{2}}{2}\phi^{2}.
\end{equation}
The equivalent first order action principle is now
\begin{equation}
  \label{eq:151}
  S_H[\phi,\pi^+,\lambda^+]=\int dx^+\int_{-L_-/2}^{{L_{-}/2}}dx^-[\pi^+\partial_+\phi-\frac{m^2}{2} \phi^2-\lambda^+(\pi^+-\partial_-\phi)].
\end{equation}

Conservation in time of the constraints $\{g^{+},H+G[\lambda^{+}]\}\approx 0$ now imposes
\begin{equation}
  \label{eq:149}
  \partial_{-}\lambda^{+}=-\frac{m^{2}}{2}\phi.
\end{equation}

In the following, we assume all fields to be periodic in $x^-$ and decompose, as
in the massless case, into a zero mode that is constant on the $x^+$ front and
only depends on $x^+$, and a field without zero mode.

It then follows that there is a single secondary constraint
that sets to zero the zero mode of $\phi$ as defined in \eqref{eq:41},
\begin{equation}
  \label{eq:147}
  \bar\phi_{{+}}\approx 0.
\end{equation}
Furthermore, it follows from \eqref{eq:149} that $\tilde \lambda^+_+$ is uniquely fixed
by $\tilde\phi_+$,
\begin{equation}
  \label{eq:141}
  \tilde \lambda^+_+(\tilde\phi_+)=-\frac{m^2}{2}\partial_-^{-1}\tilde \phi_+,
\end{equation}
Conservation in time of the secondary constraint,
$\{\bar\phi_{{+}},H+G[\lambda^{+}]\}\approx 0$, then requires the zero mode of
the Lagrange multiplier to vanish, $\bar\lambda^{+}_{+}=0$. All the
constraints, the primary ones $\tilde g^{+}_{+},\bar g^{+}_{+}=\bar\pi^+_+$ in \eqref{eq:42}
together with the secondary one $\bar\phi_{+}$ are second class. Note that the
absence of a first class constraint is consistent with the absence of a shift
symmetry. The total, first class, Hamiltonian is
\begin{equation}
  \label{eq:148}
  H'=H+G^{+}[\tilde\lambda^{+}_{{+}}(\tilde \phi_{+})].
\end{equation}

When imposing all second class constraints strongly and working with Dirac
brackets, the zero modes $\bar\phi_{+},\bar\pi^{+}_{+}$ are eliminated from the theory.
The only variables that remain are $\tilde\phi(x^{-})$ with Dirac brackets given in
the first equation of \eqref{Diracbrackets}. The Hamiltonian reduces to the one
given in \eqref{eq:144} and the general solution to the evolution equations
\begin{equation}
  \label{eq:6}
  \partial_{+}\tilde \phi_{+}(x^-)=\{\tilde\phi_{+}(x^-),H\}^{*}_{+}=m^2\int^{L_-/2}_{-L_-/2} dy^-
  [-\frac 14 \varepsilon(x^--y^-)+
  \frac{x^--y^-}{2L_-}]\tilde\phi_+(y^-), 
\end{equation}
is, for periodic boundary conditions, entirely determined by the free data
$\tilde\phi(c^{+},x^{-})$. It may be represented formally through the series
\begin{equation}
  \label{eq:145}
  \tilde\phi_{+}(x^{+},x^{-})=e^{-(x^{+}-c^{+})\{H,\cdot\}^*_{+}}\tilde\phi_{+}(c^{+},x^{-}).
\end{equation}
It thus follows that, with periodic boundary conditions in $x^-$, the free data
in the massive case can be prescribed on a single null hyperplane.
 
\section{Lessons and outlook}
\label{sec:conclusion}

The main lessons for a consistent quantization of fields on null hyperplanes
that can been drawn from this detailed investigation of the simplest example
are the following:
\begin{enumerate}[label=(\roman*)]
  
\item The Dirac algorithm systematically determines the free data not only of
  the Cauchy problem but also of the characteristic initial value problem.
  
\item The algorithm is very different in the massless and massive case. In the
  former, there are primary first class constraints and data is needed on both
  of the intersecting null planes. They are absent in the latter case and data
  on a single null surface is enough to fully determine the general solution.

  \item The first class constraint in the front form of dynamics in the massless
        case does not generate a gauge symmetry but rather an infinite number of
        global symmetries. The associated Lagrange multiplier corresponds to a
        canonical momentum for time evolution along, rather than off, the front
        and encodes information on physical degrees of freedom.
  
  \item As in the instant form of dynamics, the problem has to be supplemented
        by boundary conditions. The simplest one consists in putting the system
        in a box with periodic boundary conditions. This is where discrete
        light-cone quantization comes in. Equivalence with quantization in
        instant form on a time-like cylinder requires entangled periodic
        boundary conditions in the intersecting null directions. The free
        particle in the spectrum does not satisfy the periodic boundary
        conditions used in discrete light-cone quantization. This is where
        improvement terms of Regge-Teitelboim/ Gibbons-Hawking type become
        relevant.

\item When trying to describe on a single null hyperplane the full system as
  understood from the viewpoint of the instant form, one misses part of the
  canonical structure. The generators of the left chiral shifts and conformal
  transformations vanish on-shell and do not provide a canonical realization of
  the left chiral half of the conformal algebra.

\item A formulation on intersecting double fronts is required. The missing part of
  the symplectic structure is determined by equal time Poisson brackets on the
  other front. The left chiral shifts and the left chiral half of the conformal
  symmetries are entirely generated by non-vanishing charges on the other front
  that form a canonical realization of the algebra. Suitable matching conditions
  become relevant in order not to over-count when re-assembling the descriptions
  on the individual fronts.
  
  \item The problem of quantization of a free massless boson on intersecting
        null lines turns out to be closely related to the formulation of chiral
        $p$-forms in terms of non chiral ones (see e.g.~\cite{Bekaert:1998yp}
        and references therein) in the simplest case with $p=0$. In particular,
        the coupling of the chiral sectors with the topological zero mode sector
        in the Noether charges for the conformal symmetries is particularly
        instructive.
  
\item In order to get the correct Dirac/ Peierls brackets that canonically generate
  the transformations, a covariant first order formulation with suitable
  auxiliary fields can be useful.
\end{enumerate}

It should be quite instructive to work out the details on how to deal with
electromagnetism, Yang-Mills and gravitational theories and their intrinsic
gauge invariances from the viewpoint of the canonical approach on intersecting
double fronts, while keeping both physical and unphysical degrees of freedom. We
plan to address these questions elsewhere.

\section*{Acknowledgments}
\label{sec:acknowledgements}

\addcontentsline{toc}{section}{Acknowledgments}

G.B.~is grateful to M.~Bonte, D.~Anninos, M.~Henneaux and H.A.~Gonz\'alez for
useful discussions. This work is supported by the F.R.S.-FNRS Belgium through
convention FRFC PDR T.1025.14, and convention IISN 4.4514.08. W.T.~is supported
by a PhD grant from the China Scholarship Council. The work of S.M. is supported by the LABEX Lyon Institute of Origins (ANR-10-LABX-0066) within the Plan France 2030 of the French government operated by the National Research Agency (ANR). G.B. and S.M. are grateful to MITP for hospitality during the programs ``Higher Structures, Gravity and
Fields'' and ``Thermalization in Conformal Field Theories'' and to the
organizers of the 3rd Carroll workshop in Thessaloniki.

\appendix

\section{Instant and space forms}
\label{sec:instant-form-}

\subsection{Covariant Hamiltonian formulation}
\label{sec:position-space}

In two-dimensional Minkowski spacetime with metric
$ds^2=(dx^0)^2-(dx^1)^2=\eta_{\mu\nu}dx^{\mu}dx^{\nu}$, the Lagrangian action for a free
massless scalar field is:
\begin{equation}
  \label{eq:3a}
  S[\phi]=\frac 12 \int dx^0dx^1\,\mathcal L,\quad \mathcal L=
  \partial_\mu\phi\partial^\mu\phi.
\end{equation}
Infinitesimal conformal symmetries acts as
\begin{equation}
  \label{eq:165}
  \delta_\xi\phi=\xi^\rho\partial_\rho\phi,\quad \partial_\mu\xi_\nu+\partial_\nu\xi_\mu=\partial_\rho\xi^\rho\eta_{\mu\nu},
\end{equation}
with
\begin{equation}
  \label{eq:176}
  \frac{\delta\mathcal L}{\delta\phi}\delta_\xi\phi+\partial_\mu(T^\mu_\nu \xi^\nu)=0,\quad T_{\mu\nu}=\partial_\mu\phi\partial_\nu\phi-\frac 12 \eta_{\mu\nu}\partial_\rho\phi\partial^\rho\phi.
\end{equation}

The associated first order action in instant form is
\begin{equation}
  \label{eq:56}
  S_H[\phi,\pi^0]=\int dx^0dx^1\,\mathcal{L}_H^0,\quad \mathcal{L}_H^0=\pi^0\partial_0 \phi-\mathcal H^0,\quad \mathcal H^0=\frac 12((\pi^0)^2+(\partial_1\phi)^2).
\end{equation}

In order to better compare to the light front analysis, one may introduce an
additional auxiliary field $\pi^1$ and work with the equivalent ``covariant''
first order action
\begin{equation}
  \label{eq:125}
  S'_H[\phi,\pi^\mu]=\int dx^0dx^1\,\mathcal L'_H,\quad \mathcal L'_H=\pi^{\mu}\partial_{\mu} \phi-\frac 12 \pi^{\mu}\pi_{\mu},
\end{equation}
with equations of motion
\begin{equation}
  \label{eq:128}
  \partial_\rho\phi-\pi_\rho=0,\quad\partial_\mu\pi^\mu=0.
\end{equation}

The full Hamiltonian treatment of this system then leads to
\begin{multline}
  \label{eq:127}
  S''_H[\phi,\pi^\mu,\pi_{\pi^1},\lambda^1,\lambda^2]=\int dx^0dx^1\,[\pi^0\partial_0
  \phi+\pi_{\pi^1}\partial_0\pi^1+\pi^1\partial_1\phi-\frac 12(\pi^0)^2+\frac 12
  (\pi^1)^2\\
-\lambda^1\pi_{\pi^1}-\lambda^2(\pi^1+\partial_1\phi)],
\end{multline}
with two second class constraints and associated Dirac
brackets, $\{\pi_{\pi^1}(x^1),\cdot\}^*_0=0$ together with
\begin{equation}
  \label{eq:169}
  \begin{split}
    &  \{\phi(x^1),\pi^0(y^1)\}^*_0=\delta(x^1,y^1),\quad \{\phi(x^1),\pi^1(y^1)\}^*_0=0,\\
    &\{\pi^1(x^1),\pi^0(y^1)\}^*_0=-\delta'(x^1,y^1)=\{\pi^0(x^1),\pi^1(y^1)\}^*_0.
  \end{split}
\end{equation}
The Hamiltonian is given by
\begin{equation}
  \label{eq:185}
  H'^0=\int dx^1 [\frac 12 (\pi^0)^2-\frac 12 (\pi^1)^2-\pi^1\partial_1\phi].
\end{equation}
Imposing the second class constraints strongly
then leads one back to the theory defined by \eqref{eq:56}.

One may now also easily work out what happens if one sets up the Hamiltonian
formalism at a constant $x^1$, for instance $x^1=0$, which is a time-like
surface, so that $x^1$ is the evolution parameter. For want of a better name,
this might be called the space form, which would be relevant in holographic
settings of AdS type. The canonical variables are $\phi(x^0),\pi^1(x^0)$, the
auxiliary field is $\pi^0$, which is on-shell given by $\pi^0=\partial_0\phi$,
and forms together with $\pi_{\pi_0}$ second class constraints. The associated
Dirac brackets are $\{\pi_{\pi^0},\cdot\}^*_1=0$ and
\begin{equation}
  \label{eq:169a}
  \begin{split}
    &  \{\phi(x^0),\pi^1(y^0)\}^*_1=\delta(x^0,y^0),\quad \{\phi(x^0),\pi^0(y^0)\}^*_1=0,\\
    &\{\pi^0(x^0),\pi^1(y^0)\}^*_1=\delta'(x^0,y^0)=\{\pi^1(x^0),\pi^0(y^0)\}^*_1.
  \end{split}
\end{equation}
The Hamiltonian is given by
\begin{equation}
  \label{eq:184}
  H'^1=\int dx^0 [\frac 12 (\pi^0)^2-\frac 12 (\pi^1)^2-\pi^0\partial_0\phi],
\end{equation}
and, after imposing the constraints strongly, reduces to
\begin{equation}
  \label{eq:186}
  H^1=-\frac 12 \int dx^0 [ (\pi^1)^2+ (\partial_0\phi)^2]. 
\end{equation}

\subsection{Covariant Hamiltonian representation of shift and
  conformal symmetries}
\label{sec:conf-transf}

For the shift symmetries, we have
$\partial_0\epsilon^{\pm}=\pm\partial_1\epsilon^\pm$. The transformations
$\delta_{\epsilon^{\pm}}\phi=\epsilon^\pm$ are extended to $\pi^0$ as
$\delta_{\epsilon^\pm}\pi^0=\partial_0\epsilon^\pm$. The Noether
equations become
\begin{equation}
  \label{eq:197}
  \begin{split}
    & \delta_{\epsilon^\pm}\phi\frac{\delta \mathcal{L}_H^0}{\delta \phi}+\delta_{\epsilon^\pm}\pi^0\frac{\delta \mathcal{L}_H^0}{\delta \phi}+\partial_\mu j^\mu_{\epsilon^\pm}=0,\\
    & j^0_{\epsilon^\pm}=\pi^0\epsilon^\pm-\phi\partial_0\epsilon^\pm,\quad j^1_{\epsilon^\pm}=-\partial_1\phi\epsilon^\pm\pm\phi \partial_0\epsilon^\pm. 
  \end{split}
\end{equation}
The associated Noether charges that generate these transformations in the Poisson bracket
are $Q^0_{\epsilon^\pm}=\int dx^1 j^0_{\epsilon^\pm}$. 

Infinitesimal conformal transformations are parametrized by the vectors
$\xi^\mu(x^\nu)$, with \eqref{eq:165} equivalent to
$\partial_0\xi^0=\partial_1\xi^1$, $\partial_0\xi^1=\partial_1\xi^0$. If
$\mathcal P^0=\pi^0\partial_1\phi$, the Noether equation adapted to the
Hamiltonian formalism in instant form may be written as
\begin{equation}
  \label{eq:160}
  \delta^{0}_\xi\phi\frac{\delta \mathcal{L}_H^0}{\delta \phi}+\delta^{0}_\xi\pi^0\frac{\delta \mathcal{L}_H^0}{\delta \phi}+\partial_\mu j^\mu_\xi=0,
\end{equation}
where
\begin{equation}
  \label{eq:161}
  \begin{split}
    & \delta^{0}_\xi \phi=\xi^0\pi^0+\xi^1\partial_1\phi,\quad \delta^{0}_\xi\pi^0=\partial_1(\xi^0\partial_1\phi+\xi^1\pi^0),\\
    & j^0_\xi=\xi^0\mathcal H^0+\xi^1\mathcal P^0,\quad
      j^1_\xi=-\xi^0\mathcal P^0-\xi^1\mathcal H^0-(\xi^0\partial_1\phi+\xi^1\pi^0)(\partial_0\phi-\pi^0).
  \end{split}
\end{equation}
The associated Noether charges $Q^0_\xi=\int dx^1 j^0_\xi$ generate the
transformations through the Poisson bracket and realize the conformal
algebra,
\begin{equation}
  \label{eq:164}
\delta_\xi^0\phi=\{\phi,Q^0_\xi\}_0,\quad \delta_\xi^0\pi^0=\{\pi^0,Q^0_\xi\}_0,\quad \{Q^0_{\xi_1},Q^0_{\xi_2}\}_0=Q^0_{[\xi_1,\xi_2]}.
\end{equation}
In particular, the Hamiltonian evolution equations in $x^0$ are recovered
from the first two equations by choosing $\xi^0=1$, $\xi^1=0$.

In the covariant first order formulation, the Noether equation for shift 
symmetries is extended to the auxiliary field $\pi^1$ as
\begin{equation}
  \label{eq:177a}
  \begin{split}
    &\delta_{\epsilon^\pm}\phi\frac{\delta \mathcal L'_H}{\delta\phi}+\delta_{\epsilon^\pm}\pi^\rho\frac{\delta \mathcal L'_H}{\delta\pi^\rho}+\partial_\mu j^\mu_{\epsilon^\pm}=0,\\
    & \delta_{\epsilon^\pm}\phi=\epsilon^\pm,\quad \delta_{\epsilon^\pm}\pi^\mu=\partial^\mu\epsilon^\pm,\quad  j^\mu_{\epsilon^\pm}=\pi^\mu\epsilon^\pm-\phi\partial^\mu\epsilon^\pm.
  \end{split}
\end{equation}
The associated Noether $Q^0_{\epsilon^\pm}=\int dx^1 j^0_{\epsilon^\pm}$ generate
these transformations in the Dirac bracket.

For conformal symmetries, we have
\begin{equation}
  \label{eq:177}
  \begin{split}
    &\delta^0_\xi\phi\frac{\delta \mathcal L'_H}{\delta\phi}+\delta^0_\xi\pi^\rho\frac{\delta \mathcal L'_H}{\delta\pi^\rho}+\partial_\mu \tilde j^\mu_\xi=0,\\
    & \delta^{0}_\xi\pi^1=-\partial_1\delta^0_\xi\phi,\quad \tilde j^0_\xi=j^0_\xi,\quad \tilde j^1_\xi=j^1_\xi+\delta^0_\xi\phi\frac{\delta \mathcal L'_H}{\delta\pi^1}.
  \end{split}
\end{equation}
The Noether charges $Q^0_\xi$ generate the conformal transformations through the Dirac
bracket and realize the conformal algebra
\begin{equation}
  \label{eq:178}
  \delta_\xi^0\phi=\{\phi,Q^0_\xi\}_0^*,\quad \delta_\xi^0\pi^\rho=\{\pi^\rho,Q^0_\xi\}_0^*,\quad \{Q^0_{\xi_1},Q^0_{\xi_2}\}^*_0=Q^0_{[\xi_1,\xi_2]}.
\end{equation}
This computation relies on the assumption that one may integrate by parts in
space.

A more covariant way to write Noether's theorem for conformal symmetries in
first order form is
\begin{equation}
  \label{eq:168}
  \begin{split}
    &\delta'_\xi\phi\frac{\delta \mathcal L'_H}{\delta\phi}+\delta'_\xi\pi^\rho\frac{\delta \mathcal L'_H}{\delta\pi^\rho}+\partial_\mu j'^\mu_\xi=0,\quad
      \delta'_\xi \phi=\xi^\rho\pi_\rho,\quad \delta'_\xi\pi^\rho=\partial_\mu(\xi^\mu\pi^\rho-\xi^\rho\pi^\mu),\ \\
    &j'^\mu_\xi=\pi^\mu\xi^\rho\pi_\rho-\frac 12 \xi^\mu\pi^\rho\pi_\rho-(\xi^\mu\pi^\rho-\xi^\rho\pi^\mu)\frac{\delta \mathcal L'_H}{\delta\pi^\rho}.
  \end{split}
\end{equation}
The associated Noether charge is $Q'^0_\xi=\int dx^1 j'^0_\xi$ with
\begin{equation}
  \label{eq:181}
  j'^0_\xi=\pi^0\xi^\rho\pi_\rho-\frac 12 \xi^0\pi^\rho\pi_\rho-(\xi^0\pi^1-\pi^0\xi^1)(\pi^1+\partial_1\phi),
\end{equation}
and
\begin{equation}
  \label{eq:170}
  Q'^0_\xi=Q^0_\xi-\frac 12\int dx^1 \xi^0(\pi^1+\partial_1\phi)^2.
\end{equation}
It generates the $\delta^0_\xi$ transformations, and the $\delta'_\xi$
transformations on-shell,
\begin{equation}
  \label{eq:171}
  \{\phi,Q'^0_\xi\}^*_0=\delta^{0}_{\xi}\phi\approx \delta'_\xi\phi,\ \{\pi^0,Q'^0_\xi\}^*_0=\delta^{0}_\xi\pi^0\approx \delta'_\xi\pi^0,\
  \{\pi^1,Q'^0_\xi\}^*_0=\delta^{0}_{\xi}\pi^1\approx\delta'_\xi\pi^1,
\end{equation}
where the two first relations only require the second class constraint
$\pi^1=-\partial_1\phi$, while the last relation needs all equations of motion.
Furthermore, the Dirac bracket algebra of the Noether charges represents the
conformal algebra on the second class constraint surface,
\begin{equation}
  \label{eq:172}
  \{Q'^0_{\xi_1},Q'^0_{\xi_2}\}^*_0=Q^0_{[\xi_1,\xi_2]}=Q'^0_{[\xi_1,\xi_2]}+\frac 12 \int dx^1 [\xi_1,\xi_2]^0 (\pi^1+\partial_1\phi)^2.
\end{equation}
Finally, note that in terms of the standard Poisson bracket, if one denotes by
$\delta^{0,\pi^1}_\xi$ a variation that acts only on $\pi^1$ according to
$\delta^{0,\pi^1}_\xi\pi^1=\delta^0_\xi\pi^1=-\partial_1(\delta^0_\xi\phi)$, one may write
the Dirac bracket of the Noether charges in terms of a ``modified'' bracket as 
\begin{multline}
  \label{eq:175}
  \{Q'^0_{\xi_1},Q'^0_{\xi_2}\}^*_0=\{Q'^0_{\xi_1},Q'^0_{\xi_2}\}_0\\+\int dx\int dy \big[\frac{\delta
    Q'^0_{\xi_1}}{\delta\pi^0(x)}\{\pi^0(x),\pi^1(y)\}^*_0\frac{\delta
    Q'^0_{\xi_2}}{\delta\pi^1(y)}
+\frac{\delta
  Q'^0_{\xi_1}}{\delta\pi^1(x)}\{\pi^1(x),\pi^0(y)\}^*_0\frac{\delta
  Q'^0_{\xi_2}}{\delta\pi^0(y)}
  \big]\\
  =\{Q'^0_{\xi_1},Q'^0_{\xi_2}\}_0-\delta^{0,\pi^1}_{\xi_1}Q'^0_{\xi_2}+\delta^{0,\pi^1}_{\xi_2}Q'^0_{\xi_1}.
\end{multline}
In other words, the modified bracket arises here when one works with the standard
Poisson bracket and one forgets that the transformation law of the auxiliary field
$\pi^1$ is canonically generated by the appropriate Dirac bracket.

In space form, the Noether charges for shift symmetries are
$Q^1_{\epsilon^\pm}=\int dx^0 j^1_{\epsilon^\pm}$ and generate the
transformations in the associated Dirac brackets. For conformal symmetries, they
are $Q'^1_{\xi}=\int dx^0 j'^1_\xi$ where
\begin{equation}
  \label{eq:180}
  j'^1_\xi=\pi^1\xi^\rho\pi_\rho-\frac 12 \xi^1\pi^\rho\pi_\rho-(\xi^1\pi^0-\xi^0\pi^1)\frac{\delta \mathcal L'_H}{\delta \pi^0}.
\end{equation}
They may be written as
\begin{equation}
  \label{eq:170a}
  Q'^1_\xi=Q^1_\xi+\frac 12\int dx^0 \xi^1(\pi^0-\partial_0\phi)^2.
\end{equation}
where
\begin{equation}
  \label{eq:183}
  Q^1_\xi=\int dx^0[\xi^1\mathcal H^1+\xi^0\mathcal P^1],\quad \mathcal H^1=-\frac 12 [ (\pi^1)^2+(\partial_0\phi)^2],\quad \mathcal P^1=\pi^1\partial_0\phi,
\end{equation}
are the charges in the reduced theory without the auxiliary field $\pi^0$. The
conformal transformations generated in the Dirac bracket are
\begin{equation}
  \label{eq:179}
  \begin{split}
    & \delta^1_\xi\phi=\{\phi,Q'^1_{\xi}\}^*_1=-\xi^1\pi^1+\xi^0\partial_0\phi,\quad \delta^1_\xi\pi^1=\{\pi^1,Q'^1_{\xi}\}^*_1=\partial_0(-\xi^1\partial_0\phi+\xi^0\pi^1),\\
    &\delta^1_\xi\pi^0=\{\pi^0,Q'^1_{\xi}\}^*_1=\partial_0(\delta^1_\xi\phi),
  \end{split}
\end{equation}
and
\begin{equation}
  \label{eq:182}
  \{Q'^1_{\xi_1},Q'^1_{\xi_2}\}^*_1=Q^1_{[\xi_1,\xi_2]}=Q'^1_{[\xi_1,\xi_2]}-\frac 12 \int dx^0 [\xi_1,\xi_2]^1 (\pi^0-\partial_0\phi)^2.
\end{equation}
Note that these computations rely on the assumption that one may integrate by
parts in time.

\subsection{Decomposition on a time-like cylinder}
\label{sec:decomposition-cylinder}

The analysis starts by following the discussion in \cite{Henneaux1989} on how to
perform an off-shell split of a free massless boson in two dimensions in terms
of left and right moving chiral fields and an $x^1$ zero mode.

We consider the system \eqref{eq:56} on a spatial circle of length $L$. In other
words, the points $(x^0,x^1)\sim (x^0,x^1+L)$ are identified and all fields
satisfy periodic boundary conditions in $x^1$.

First, one separates the $x^1$ zero mode from the rest of
the fields,
\begin{equation}
  \label{eq:63}
  \begin{split}
    \phi(x^0,x^1)=\bar \phi_0(x^0)+\tilde \phi_0(x^0,x^1),\quad\pi^0(x^0,x^1)
    =\frac{1}{L}\bar \pi^0_0(x^0)+\tilde\pi^0_0(x^0,x^1),\\
    \bar \phi_0=\frac{1}{L}\int_{-L/2}^{L/2}dx^1\, \phi,\quad \bar\pi^0_0=\int^{L/2}_{-L/2}dx^1\, \pi^0,\quad
    \int_{-L/2}^{L/2} dx^1\, \tilde \phi_0=0=\int_{-L/2}^{L/2}dx^1\,\tilde\pi_0^0.
  \end{split}
\end{equation}
In terms of this decomposition, the action \eqref{eq:56} splits as
$S_P[\bar\phi_0,\bar\pi^0_0]+S_H[\tilde\phi_0,\tilde \pi^0_0]$ with
\begin{equation}
  \label{eq:126}
  S^P[\bar\phi_0,\bar\pi^0_0]=\int dx^0[\bar\pi^0_0\partial_0\bar\phi_0-\frac{1}{2L} (\bar\pi^0_0)^2].
\end{equation}
The equations of motion become
\begin{equation}
  \label{eq:76}
    \partial_0 \bar\phi_0=\frac{1}{L}\bar\pi^0_0,\
    \partial_0\bar\pi^0_0=0,\ \tilde \pi^0_0=\partial_0\tilde\phi_0,\ \partial_0 \tilde\pi^0_0-\partial_1^2\tilde \phi_0=0,
  \end{equation}
and imply the wave equation $\partial_0^2\tilde\phi_0=\partial_1^2\tilde\phi_0$.

The non-vanishing equal $x^0$
  time Poisson brackets are
  \begin{equation}
    \label{eq:81}
    \{\phi(x^1),\pi^0(y^1)\}_0=\delta^P(x^1,y^1),\quad \delta^P(x^1,y^1)=\frac{1}{L}\sum_{n\in\mathbb Z}e^{ik(x^1-y^1)}.
  \end{equation}
  After the split, they become
\begin{equation}
  \label{eq:88}
  \{\tilde \phi_0(x^1),\tilde \pi^0_0(y^1)\}_0=\delta^P(x^1,y^1)-\frac{1}{L},\quad
    \{\bar\phi_0,\bar\pi^0_0\}_0=1.
\end{equation}
Then the chiral fields are defined as
\begin{equation}
  \label{eq:77}
  \phi^L=\frac 12 (\tilde\phi_0+ \partial_1^{-1}\tilde \pi^0_0),\quad \phi^R=
  \frac 12 (\tilde\phi_0  -\partial_1^{-1}\tilde \pi^0_0).
\end{equation}
Note that on fields without a zero mode,
there is a precise definition of $\partial_1^{-1}$ that does not contain a zero mode, 
\begin{equation}
  \label{eq:206}
  \partial^{-1}_1\tilde\pi^0_0(x^0,x^1)=\int^{x^1}_{0}dy^1\,\tilde \pi^0_0(x^0,y^1)
  -\frac{1}{L}\int_{-L/2}^{L/2}dy^1\int_0^{y^1}dz^1\tilde \pi^0_0(x^0,z^1).
\end{equation}
Alternatively, for periodic fields in $x^1$, one integrates the zero mode free
Fourier series term by term in such a way that the result is again a zero
mode free Fourier series.

This definition guarantees (i) that the chiral fields do not involve a zero
mode, (ii) a simple invertible off-shell relation for the fields and their
momenta in terms of the chiral fields
\begin{equation}
  \label{eq:92}
  \tilde\phi_0=\phi^L+\phi^R,\quad \tilde\pi^0_0=\partial_1\phi^{L}-\partial_1\phi^R,
\end{equation}
and, (iii) that the equations of motion
\eqref{eq:76} imply that the chiral fields are left and right movers,
\begin{equation}
  \label{eq:78}
  \partial_0\phi^L=\partial_1\phi^L,\quad \partial_0\phi^R=-\partial_1\phi^R.
\end{equation}
The general solution to the equations of motion is
\begin{equation}
  \label{eq:79}
  \phi^L=\phi^L(x^+),\quad \phi^R=\phi^R(x^-),\quad \bar\pi_0^0=\bar\pi^0_0(c_0),\quad
  \bar\phi_0=\bar\phi_0(c^0)+\frac{x^0}{L}\bar\pi^0_0(c_0).
\end{equation}
Up to a boundary term in time, there is an off-shell split of the action
principle into left and right moving chiral bosons,
\begin{equation}
  \label{eq:191}
  S_H[\tilde\phi_0,\tilde\pi^0_0]=S^L[\phi^L]+S^R[\phi^R] +S^B,
\end{equation}
with
\begin{equation}
  \begin{split}
  \label{eq:89}
    S^L[\phi^L]&= \int_{x^0_i}^{x^0_f} dx^0\int_{-L/2}^{L/2}
    dx^1\big[\partial_1\phi^L\partial_0\phi^L-(\partial_1\phi^L)^2\big],\\S^R[\phi^R] &=\int_{x^0_i}^{x^0_f} dx^0\int_{-L/2}^{L/2}
              dx^1\big[-\partial_1\phi^R\partial_0\phi^R-(\partial_1\phi^R)^2\big],\\
        S^B&= \Big[\int_{-L/2}^{L/2} dx^1\frac 12(\partial_1\phi^L\phi^R-\phi^L\partial_1\phi^R) \Big]^{x^0_f}_{x^0_i}.
  \end{split}
\end{equation}
Note that, while action \eqref{eq:56} is invariant under parity transformations,
\begin{equation}
  \label{eq:120}
  S_H[\hat\phi,\hat \pi]=S_H[\phi,\pi],\quad
  \hat \phi(x^0,x^1)=\phi(x^0,-x^1),\ \hat \pi(x^0,x^1)=\pi(x^0,-x^1),
\end{equation}
they exchange the left and right chiral actions,
\begin{equation}
  \label{eq:121}
  \begin{split}
    & S^L[\hat \phi^L]=S^R[\phi^R],\quad S^R[\hat \phi^R]=S^L[\phi^L],\\
    & \hat \phi^L(x^0,x^1)=\phi^R(x^0,-x^1),\quad\hat\phi^R(x^0,x^1)=\phi^L(x^0,-x^1).
  \end{split}
\end{equation}

Note also that the primitive
$\partial_1^xu^P(x^1,y^1)=-\partial_1^yu^P(x^1,y^1)=\delta^P(x^1,y^1)-\frac{1}{L}$
that does not contain $x^1$ zero modes is
\begin{equation}
  \label{eq:103}
  u^P(x^1,y^1)=\sum_{n>0}\frac{1}{\pi n}\sin{k(x^1-y^1)}=\frac 12\varepsilon(x^1-y^1)-\frac{x^1-y^1}{L},\quad k=\frac{2\pi n}{L},
\end{equation}
where the last equality follows from the Fourier transforms of the sign function
$\varepsilon(x)$ and of the linear function $x$ on the interval
$[-\frac L2,\frac L2]$, and extended as periodic functions beyond,
\begin{equation}
  \label{eq:107}
  \varepsilon(x)=\sum_{n>0}\frac{2(1-\cos\pi n)}{\pi n}\sin{kx},\quad \frac{x}{L}=
  -\sum_{n>0}\frac{\cos\pi n}{\pi n}\sin{kx}.
\end{equation}
It follows that the equal $x^0$ time canonical Poisson brackets \eqref{eq:88} become
\begin{equation}
  \label{eq:91}
  \begin{split}
   & \{\phi^R(x^1),\phi^R(y^1)\}_0=\frac 14 \varepsilon(x^1-y^1)-\frac{x^1-y^1}{2L}=- \{\phi^L(x^1),\phi^L(y^1)\}_0,\\
   &  \{\phi^L(x^1),\phi^R(y^1)\}_0=0.
  \end{split}
\end{equation}

One may also introduce the auxiliary field $\pi^1$ (and its conjugate momentum),
and split off their zero modes. It then follows that $\bar\pi^1_0=0$,
$\tilde\pi^1_0+\partial_1\phi^L+\partial_1\phi^R=0$ together with the analogous spit of
$\pi_{\pi^1}$ are all second class constraints. If one decides to
keep $\bar\pi^1_0$, $\tilde\pi^1_0$, it follows that the Dirac brackets of
$\bar\pi^1_0$ with all variables vanish, while those of $\tilde\pi^1_0$ are given by
\begin{equation}
  \label{eq:106}
  \{\tilde\pi^1_0(x^1),\phi^R(y^1)\}^*_0=\frac 12 (\delta(x^1,y^1)-\frac 1L)
=-\{\tilde\pi^1_0(x^1),\phi^L(y^1)\}^*_0. 
\end{equation}

The Hamiltonian and linear momentum
\begin{equation}
  \label{eq:97}
  H=\frac{1}{2}\int_{-L/2}^{L/2}dx^1\, [(\pi^0)^2+(\partial_1\phi)^2],\quad P=-\int_{-L/2}^{L/2}dx^1\, \pi^0\partial_1\phi,
\end{equation}
split as $H=H^L+H^R+H^P$, $P=P^L+P^R$, with
\begin{equation}
  \label{eq:96}
  \begin{split}
    H^L&=\int_{-L/2}^{L/2}dx^1\, (\partial_1\phi^L)^2,\quad H^R=\int_{-L/2}^{L/2}dx^1\, (\partial_1\phi^R)^2,\quad
        H^P=\frac{1}{2L}(\bar\pi^0_0)^2,\\
    P^L&=-\int_{-L/2}^{L/2}dx^1\, (\partial_1\phi^L)^2,\quad P^R=\int_{-L/2}^{L/2}dx^1\, (\partial_1\phi^R)^2.
  \end{split}
\end{equation}

Of interest to us here is the extended partition function
\begin{equation}
  \label{eq:56b}
  Z(\beta,\alpha)={\rm Tr}\, e^{-\beta  \hat H +i\alpha \hat P}.
\end{equation}
In terms of the modular parameter $\tau=\frac{\alpha+i\beta}{L}=\tau_1+i\tau_2$,
the combination of Hamiltonian and linear momentum that is involved decomposes
into right and left chiral pieces together with a free particle contribution,
\begin{equation}
  \label{eq:98}
  -\beta H+i\alpha P=i\tau L H^R-i\bar\tau L H^L-\tau_2L H^P.
\end{equation}

\subsection{Decomposition of conformal transformations}
\label{sec:decomp-conf-transf}

It is also useful to see how the infinitesimal conformal transformations mix the
different sectors.

This leads to
\begin{equation}
  \label{eq:162}
  \begin{split}
    & \delta_\xi\bar\phi_0=\frac{\bar\pi^0_0}{L^2}\int_{-L/2}^{L/2}dx^1 \xi^0
      +\frac{1}{L}\int_{-L/2}^{L/2}dy^1\,(\xi^0\tilde\pi^0_0+\xi^1\partial_1\tilde\phi_0),\quad
    \delta_\xi\bar\pi^0_0=0,\\
    & \delta_\xi\tilde\phi_0=\xi^0\tilde\pi^0_0+\xi^1\partial_1\tilde\phi_0
      -\frac{1}{L}\int_{-L/2}^{L/2}dy^1\,(\xi^0\tilde\pi^0_0+\xi^1\partial_1\tilde\phi_0)
      +(\xi^0-\frac{1}{L}\int_{-L/2}^{L/2}dx^1 \xi^0)\frac{\bar\pi^0_0}{L},\\
    &\delta_\xi\tilde\pi^0_0=\partial_1(\xi^0\partial_1\tilde\phi_0+\xi^1\tilde\pi^0_0)
      +\frac{1}{L}\partial_1\xi^1\bar\pi^0_0.
  \end{split}
\end{equation}
These transformations are canonically generated by the Noether charge
\begin{equation}
  Q_\xi=\int_{-L/2}^{L/2}dx^1[\xi^0\tilde{\mathcal H}+\xi^1\tilde{\mathcal P}]
  +\frac{\bar\pi^0_0}{L}\int_{-L/2}^{L/2}dx^1[ \xi^0\tilde\pi^0_0+\xi^1\partial_1\tilde\phi_0]
  +\frac{(\bar\pi^0_0)^2}{2L^2}\int_{-L/2}^{L/2}dx^1 \xi^0,
\end{equation}
with $\tilde{\mathcal H}=\frac 12 [(\tilde\pi^0_0)^2+(\partial_1\tilde\phi_0)^2]$,
$\tilde{\mathcal P}=\tilde\pi^0_0\partial_1\tilde\phi_0$. For the transformation
not to mix the zero and non-zero mode sectors, the middle term has to vanish,
which is the case if $\partial_1\xi^\mu=0$. Because of the conformal Killing
equations \eqref{eq:165}, this requires the $\xi^\mu$ to be constants so that
the transformations are infinitesimal translations.

In the general case, when expressed in terms of the chiral
fields, the charge becomes
\begin{multline}
  \label{eq:163}
  Q_\xi={\sqrt 2}\int_{-L/2}^{L/2}dx^1[\xi^+(\partial_1\phi^L)^2+\xi^-(\partial_1\phi^R)^2]
  +\frac{\sqrt 2\bar\pi^0_0}{L}\int_{-L/2}^{L/2}dx^1[\xi^+\partial_1\phi^L-\xi^-\partial_1\phi^R]\\
  +\frac{(\bar\pi^0_0)^2}{2L^2}\int_{-L/2}^{L/2}dx^1 \xi^0.
\end{multline}
The transformation of $\phi^L,
\phi^R$, which only involve $\xi^+$ and $\xi^-$ respectively,
can then be obtained by using \eqref{eq:91}.

\subsection{Mode expansions and partition function}
\label{sec:mode-expans-part}

Let us choose initial data at $x^0=0$ and define
\begin{equation}
  \label{eq:90}
  k=\frac{2\pi n}{L}=-k_1,\quad
  k_0=|k|,\quad k_\mu x^\mu=|k|x^0-k x^1.
\end{equation}
The standard mode expansion for the on-shell
fields and momenta is
\begin{equation}
  \label{eq:80}
  \begin{split}
    \phi(x^0,x^1)=\bar\phi_0(0)+\frac{x^0}{L}\bar\pi^0_0(0)+\sum_{n\in \mathbb Z^*}
    \Big[\frac{1}{\sqrt{2|k|L}}a_{k} e^{-ik_\mu x^\mu}
    +{\rm c.c.}\Big],\\
    \pi(x^0,x^1)=\frac{1}{L}\bar \pi^0_0(0)-i\sum_{n\in \mathbb Z^*}
    \Big[\sqrt{\frac{|k|}{2L}}a_{k} e^{-ik_\mu x^\mu}-{\rm c.c.}\Big].
  \end{split}
\end{equation}
In terms of oscillators, the equal $x^0$ time canonical Poisson brackets
\eqref{eq:81} are given by
\begin{equation}
  \label{eq:82}
  \begin{split}
  \{a_k,a^*_{k'}\}_0&=-i\delta_{n,n'},\ \{a_k,a_{k'}\}_0=0= \{a^*_k,a^*_{k'}\}_0,\\\{\bar\phi_0,\bar\pi^0_0\}_0&=1,\
    \{\bar\phi_0,\bar\phi_0\}_0=0=\{\bar\pi^0_0,\bar\pi^0_0\}_0.
  \end{split}
\end{equation}

From \eqref{eq:92}, it follows that the expansions of the derivatives of the
on-shell chiral fields are
\begin{equation}
  \label{eq:93}
  \partial_1\phi^R=i\sum_{n>0}\sqrt{\frac{k}{2L}}(a_ke^{ik(x^1-x^0)}-{\rm c.c.}),\quad
  \partial_1\phi^L=-i\sum_{n>0}\sqrt{\frac{k}{2L}}(\tilde a_ke^{-i k(x^1+x^0)}-{\rm c.c.}),
\end{equation}
where $\tilde a_k=a_{-k}$. One may now integrate, which determines $\phi^R,\phi^L$
up to functions of time $f^R(x^0),f^L(x^0)$. These functions have to vanish because
the chiral fields have been defined so as not to contain an $x^1$ zero mode. Hence,
\begin{equation}
      \label{eq:94}
      \phi^R(x^-)=\sum_{n>0}\frac{1}{\sqrt{4\pi n}}(a_ke^{-i k_-x^-}+{\rm c.c.}),\quad
      \phi^L(x^+)=\sum_{n>0}\frac{1}{\sqrt{4\pi n}}(\tilde a_ke^{-i k_+ x^+}+{\rm c.c.}),
\end{equation}
where
\begin{equation}
  \label{eq:104}
  x^\pm=2^{-\frac 12}(x^0\pm x^1),\quad L_{\pm}=\frac{L}{\sqrt 2},\quad k_{\pm}=\frac{2\pi n}{L_{\pm}}.
\end{equation}
This also follows directly from the definition \eqref{eq:77}. In particular,
having extracted the $x^1$ zero mode from the chiral fields uniquely fixes the
prescription for the action of $\partial_1^{-1}$ on those fields.
As a consequence, the on-shell field may be decomposed as
\begin{equation}
  \label{eq:114}
  \phi(x^{0},x^{1})=\bar\phi_{0}(0)+\frac{x^{0}}{L}\bar\pi^{0}_{0}(0)+\phi^{R}(x^{-})+\phi^{L}(x^{+}).
\end{equation}

One then recovers from the mode expansions and \eqref{eq:82} the equal $x^0$ time
Poisson brackets of the chiral fields,
\begin{equation}
  \label{eq:105}
  \begin{split}
    \{\phi^R(x^1),\phi^R(y^1)\}_0&=\sum_{n>0}\frac{\sin{k(x^1-y^1)}}{2\pi n}=-
    \{\phi^L(x^1),\phi^L(y^1)\}_0,\\ \{\phi^L(x^1),\phi^R(y^1)\}_0&=0,
  \end{split}
\end{equation}
in agreement with \eqref{eq:91}.
The chiral Hamiltonians are
\begin{equation}
   \label{eq:83}
   H^R=\frac{1}{2}
   \sum_{n>0}k(a^*_ka_k+a_ka^*_k),\quad H^L=\frac{1}{2}\sum_{n>0}k(\tilde a^*_k\tilde a_k+\tilde a_k\tilde a^*_k).
 \end{equation}
 When defining the divergent zero point Casimir energy through zeta function
 regularization, the combination of the quantum chiral Hamiltonians that appears
 in the extended partition function is
\begin{equation}
  \label{eq:99}
  i\tau L \hat H^R=2\pi i \tau (\sum_{n>0} n \hat a^\dagger_k\hat a_k-\frac{1}{24}),\quad
  -i\bar \tau \hat H^L=-2\pi i \bar \tau (\sum_{n>0} n \hat{\tilde a}^\dagger_k\hat{\tilde a}_k-\frac{1}{24}).
\end{equation}
The modular invariant extended partition function then factorizes into right
chiral, left chiral, and particle contributions,
\begin{equation}
  \label{eq:95}
  Z(\tau,\bar \tau)=Z^R(\tau)Z^L(\bar \tau)Z^P(\tau,\bar\tau).
\end{equation}
The former are given in terms of the Dedekind $\eta$ function
\begin{equation}
  \label{eq:101}
  \eta(q)=q^{\frac{1}{24}}\prod_{n\in \mathbb N^*}(1-q^n),\quad q(\tau)=e^{2\pi i \tau},
\end{equation}
as
\begin{equation}
  \label{eq:100}
  Z^R(\tau)=\frac{1}{\eta(q(\tau))},\quad Z^L(\bar\tau)=\frac{1}{\widebar{\eta(q(\tau)}}.
\end{equation}
Finally, after dropping a standard divergence while keeping the finite part, the particle
contribution is given by
\begin{equation}
  \label{eq:102}
  Z^P(\tau,\bar\tau)=\frac{1}{\sqrt{\tau_2}},
\end{equation}
so as to achieve the modular invariant result \eqref{eq:63a}. 

\section{Peierls bracket for a free massless boson in 1+1 dimensions}
\label{sec:peierls-bracket}

The relevant object for the Peierls bracket is the difference of the advanced
minus the retarded Green's function,
\begin{equation}
  \label{eq:192}
  \tilde G(x^0,x^1)=G^+(x^0,x^1)-G^{-}(x^0,x^1),
\end{equation}
which is minus the Pauli-Jordan commutation function
$\tilde G(x^0,x^1)=-D_0(x^0,x^1)$. For the wave operator associated to the free
massless boson in 2 dimensions (see e.g.~\cite{bogolubov2012general}, chapter
11), it satisfies
\begin{equation}
  \label{eq:189}
  (\partial_0^2-\partial_1^2)\tilde G(x^0,x^1)=0,\quad \tilde G(0,x^1)=0,\quad \partial_0 \tilde G(x^0,x^1)|_{x^0=0}=-\delta(x^1),
\end{equation}
and is explicitly given by
\begin{align}
  \label{eq:190}
   \tilde G(x^0,x^1)&=-\int_{-\infty}^{+\infty}dk^1\frac{1}{4\pi k^1}[\sin{k_1(x^0+x^1)}+\sin{k^1(x^0-x^1)}]\\
                  &=-\frac{1}{4}[\varepsilon(x^0+x^1)+\varepsilon(x^0-x^1)]=-\frac 12 \varepsilon(x^0)\theta(x_\mu x^\mu),\label{eq:190b}
                    \\
  &=\int_{-\infty}^{+\infty}dk^0\int_{-\infty}^{+\infty}dk^1\frac{1}{2\pi i}e^{-ik_\mu x^\mu}\delta(k_\mu k^\mu)\varepsilon(k^0).
\end{align}
where the following relations are useful
\begin{equation}
  \label{eq:123}
  \begin{split}
    &  \delta(k_\mu k^\mu)=\frac{1}{2|k^1|}\delta(k^0-k^1)+\frac{1}{2|k^1|}\delta(k^0+k^1),\\
    &  \delta(k_\mu k^\mu)=\frac{1}{2|k^-|}\delta(k^+)+\frac{1}{2|k^+|}\delta(k^-),\\
    & \int^{+\infty}_{-\infty}\frac{dk}{2\pi} e^{ikx}=\delta(x),\quad \int^{+\infty}_{-\infty}\frac{dk}{2\pi} \mathcal P(\frac{2}{ik})e^{ikx}
      =\int^{+\infty}_{-\infty}\frac{dk}{\pi} \frac{\sin{k x}}{k}=\epsilon(x),\\
    &  \int^{+\infty}_{-\infty}{dk} \frac{e^{-ikx}}{k+i\epsilon}=-2\pi i\theta(x). 
  \end{split}
\end{equation}
for the manipulations.

In these terms, 
\begin{equation}
  \label{eq:193}
  \{\phi(x^0,x^1),\phi(y^0,y^1)\}=\tilde G(x^0-y^0,x^1-y^1),
\end{equation}
and also
\begin{equation}
  \label{eq:194}
  \{\phi(x^0,x^1),\phi(x^0,y^1)\}=0,\quad\{\phi(x^0,y^1),\partial_0\phi(x^0,y^1)\}=\delta(x^1,y^1).
\end{equation}
Since the Peierls bracket applied to equations of motion vanishes, it is well
defined on-shell and it follows that
\begin{equation}
  \label{eq:195}
  \begin{split}
    & \{\phi(x^0,x^1),\pi^0(x^0,y^1)\}=\delta(x^1,y^1),\quad\{\phi(x^0,x^1),\pi^1(x^0,y^1)\}=0,\\
    & \{\pi^1(x^0,x^1),\pi^0(x^0,y^1)\}=-\delta'(x^1,y^1).
  \end{split}
\end{equation}

\vfill
\pagebreak

\addcontentsline{toc}{section}{References}

\printbibliography

\end{document}